\documentclass[10pt,a4paper,twocolumn]{article}
\usepackage{f1000_styles}
\usepackage{url}

\usepackage[numbers]{natbib}

\usepackage{hyperref}
\usepackage{longtable}
\usepackage{booktabs}
\usepackage[utf8]{inputenc}

\newcommand{\comment}[1]{}

\newcommand{\ONOTE}[1]{}

\renewcommand\hyperlink[2]{{\bf #2}}

\newcounter{mynum}
\newcommand{\bibno}[1]{\stepcounter{mynum}\arabic{mynum}}

\graphicspath{{./HIRES-EPS/}}

\begin{document}

\title{Sketch of a novel approach to a neural model}
\author[1]{Gabriele Scheler}
\affil[1]{Carl Correns Foundation for Mathematical Biology, Mountain View, CA 94040}

\maketitle
\thispagestyle{fancy}

\begin{abstract}
\emph{There is room on the inside.} We present an
account of neuroplasticity with respect to cell-internal processing
pathways and their relation to membrane and synaptic plasticity. We
think traditional synapse-centric, weight-based models of memorization
are not sufficient or adequate to capture the complexity of
neuroplasticity. In standard accounts, we model a network of neurons
connected by adaptive transmission links. The adaptation of these
transmission links is overly simplified using short-term and long-term
potentiation/depression, assuming weight changes according to use of the
transmission link. In contrast, we propose a paradigm switch from a
synapse-centric model (each synapse learns independently, based on its
history of use) to a neuron-centric model (each neuron uses signal
selection for intracellular pathways to express plasticity at the
membrane). Each neuron has a `vertical' dimension where internal
parameters steer the external membrane- and synapse-expressed
parameters. A neural model consists of (a) expression of parameters at
the membrane, in particular dendritic synapses or spines, and axonal
boutons (b) internal parameters in the sub-membrane zone and the
cytoplasm with its protein signaling network and (c) core parameters in
the nucleus for genetic and epigenetic information. In a neuron-centric
model, each node (=neuron) in the horizontal network has its own
internal memory. Neural transmission and information storage are
separated, not automatically combined by coupling strength. There is
filtering and selection of signals for storage. Not every transmission
event leaves a trace. This represents an important conceptual advance
over synaptic weight models. We present the neuron as a self-programming
device, rather than as passively determined by ongoing input. We believe
a new approach to neural modeling is necessary, because the experimental
evidence is not well captured by traditional synapse-centric models.
Ultimately, we are interested in the possibilities of a flexible memory
system that processes external signals according to its inherent
structure. 
\end{abstract}

\section*{Keywords}
plasticity, learning, neural model, vertical integration, internal parameters, neuroplasticity, memory, synaptic plasticity

\clearpage

\thispagestyle{empty}

\subsection*{Introduction: The neuron as a system with internal and
external
parameters}\label{introduction-the-neuron-as-a-system-with-internal-and-external-parameters}

\subsubsection*{The current framework}\label{the-current-framework}

Experimental research on neurons, and as a consequence theoretical
analysis, is often divided into electrophysiology and molecular biology.
The first deals with membrane potentials, spikes and activations in
networks of neurons linked via synapses, the second deals with
intracellular signaling, genetic and epigenetic expression, regulation
of receptors, ion channels, transporters etc. In both cases,
signal-induced processes of plasticity are investigated. But data from
both types of experiments are difficult to integrate. In the first case,
we are looking mainly at induction of synaptic plasticity, with its
variants such as Hebbian plasticity, STDP, and other forms of synaptic
weight changes based on transmission between neurons. In the second case
the data are much more varied, but changes in protein abundance,
membrane protein expression and gene regulation are observed under a
variety of stimuli.

The master framework for neural plasticity that focuses on associative
synaptic plasticity, is usually expressed in the paradigm of long-term
potentiation or depression (\protect\hyperlink{ref1}{Malenka and Bear
2004}; \protect\hyperlink{ref2}{Citri and Malenka 2008};
\protect\hyperlink{ref3}{Kullmann and Lamsa 2007};
\protect\hyperlink{ref4}{Remy and Spruston 2007};
\protect\hyperlink{ref5}{Lei et al. 2003};
\protect\hyperlink{ref6}{Zhong et al. 2006}). Many interesting and
relevant criticisms and alternative suggestions have been raised (\protect\hyperlink{ref7}{Arshavsky 2021},
\protect\hyperlink{ref8}{2022}; \protect\hyperlink{ref9}{Langille and
Gallistel 2020}; \protect\hyperlink{ref10}{Trettenbrein 2016};
\protect\hyperlink{ref11}{Abraham et al. 2019};
\protect\hyperlink{ref12}{Gallistel and Matzel 2013};
\protect\hyperlink{ref13}{Han et al. 2022}). Also, complex dynamic
models of intracellular signaling (\protect\hyperlink{ref14}{Baudot et
al. 2008}; \protect\hyperlink{ref15}{Haney et al. 2010};
\protect\hyperlink{ref16}{Crampin et al. 2004}) and genetic read-out (\protect\hyperlink{ref17}{Ciliberti et al. 2007};
\protect\hyperlink{ref18}{Lee et al. 2017};
\protect\hyperlink{ref19}{Baudry et al. 2015}) exist, as well as
elaborate simulation models (\protect\hyperlink{ref20}{Scheler 2013}).
Attempts have been made to integrate kinetic modeling of protein
signaling into neuron simulators (\protect\hyperlink{ref21}{Bhalla
2002}; \protect\hyperlink{ref22}{Vayttaden and Bhalla 2004}). These
simulators were based on compartment-based dendritic simulations, and
did not challenge the dominant paradigm of synapse-centric plasticity.
They did not gain much traction. From our perspective, such simulation
models failed to capture relevant abstractions, such as the storage of
information in internal parameters, the strong selectivity of the
internal systems for memory of events, or the time-course of storage
implied by the tiered internal system.

To further illustrate this, \protect\hyperlink{f1}{Figure 1} contrasts (\textbf{A}) a conventional AMPA/NMDA based synaptic plasticity model
with ( \textbf{B}) a fuller view of external membrane and cell-internal
events in response to signals. We observe that a theory of use-dependent
synaptic plasticity requires direct transmission from synaptic AMPA/NMDA
receptor activation to the nucleus and back. The basis for this direct
transmission (``synaptic tagging'', s. below) has not been successfully
experimentally established. The neuron's neuromodulatory (NM) input and
the plasticity of its dendritic ion channels are not integrated into the
model - often they are seen as static or serving only calibration
purposes (\protect\hyperlink{ref23}{Desai 2003}). In contrast, in a
neuron-centric view, the neuron undergoes plasticity events - mostly
mediated by synaptic input events - which affect receptor binding and
(de-)sensitization, protein expression, and genetic read-out. This
implies plasticity by different time courses, integration of synaptic
events by the respective neuronal cell, and the relevance of internally
stored parameters. Because of the complexity of the molecular events to
establish plasticity, a strong selection of signals, plus accumulation
over time, or similar compounding mechanism is to be expected. In a
synapse-centric model, the signal at the synapse is the only driver of
plasticity. In a neuron-centric model, the interaction of pre- and
post-synaptic events is considered a separate and important step.

We have set up a master equation
\begin{equation}
\Delta p\text{=}f\left( I_{p},S_{p} \right) \hfill\mbox{[internal plasticity
master equation]}
\end{equation}
with \(p\) as an external membrane parameter (e.g. synaptic strength
\(w\)), \(I\) as internal state and \(S\) as external signal or stimulus
guiding the change \(\Delta\) of parameter \(p\). This means that any
learning rule is dependent on the stimulus \(S\) in the context of the
state \(I\) of the system. Compare with the Hebbian rule or other
synaptic rule (backpropagation):
\begin{equation}
w_{ij}^{t}\text{=}w_{ij}^{t - 1}\left( 1\text{+}\lambda\left( o_{i}^{t - 1} \ast o_{j}^{t - 1} \right) \right)\hfill\mbox{[Hebbian schema]}
\end{equation}
dependent on output of neuron \(i\) and neuron \(j\).
\begin{equation}
w_{ij}^{t}\text{=}w_{ij}^{t - 1}\left( 1\text{+}\lambda\left( o_{i}^{t - 1} - o'_{i} \right) \ast o_{j}^{t - 1} \right)
\hfill
\mbox{[backpropagation]}
\end{equation}
dependent on output of neurons \(i\), \(j\) and desired output
~\(o'_{i}\), where \(\left( o_{i} - o'_{i} \right)\) =
\textquotesingle error term\textquotesingle.

In these cases there is no reference to the responsitivity of the
system. However, with \(\lambda\) as the learning rate, we may link this
parameter to internal state, and adjust it independently for each
neuron. This could also be localized to each membrane site. Then this
parameter would have to be computed based on the internal signaling
system, and represents earlier inputs as well as core diversity
\begin{equation}
\Delta p\text{=}f\left( \lambda_{p},S_{p} \right)
\end{equation}
where \(\lambda_{p}\) is the internal state and
\begin{equation}
\lambda_{p}\text{=}o(ISS)
\end{equation}
In this model \(ISS\) is the dynamic signaling system, which stores and
computes internal values. Here there is only one parameter for each site
(\(\lambda_{p}\)). This does not explain the generation of all output
values over space and time by the \(ISS\), nor the capacity for multiple
internal parameters. Still it provides an easy explanation for a shift
in setting up models of neuron plasticity. The internal response
parameter \(\lambda\) here equals the current state of the signaling
system \(ISS\) at site \(p\), i.e. a time-varying, local parameter. The
master equation (Eq~(1)) provides a schema for models with
neuron-internal dynamic state memories.

\begin{figure*}[htb]
\begin{center}
{\bf A}
\includegraphics[width=0.45\textwidth]{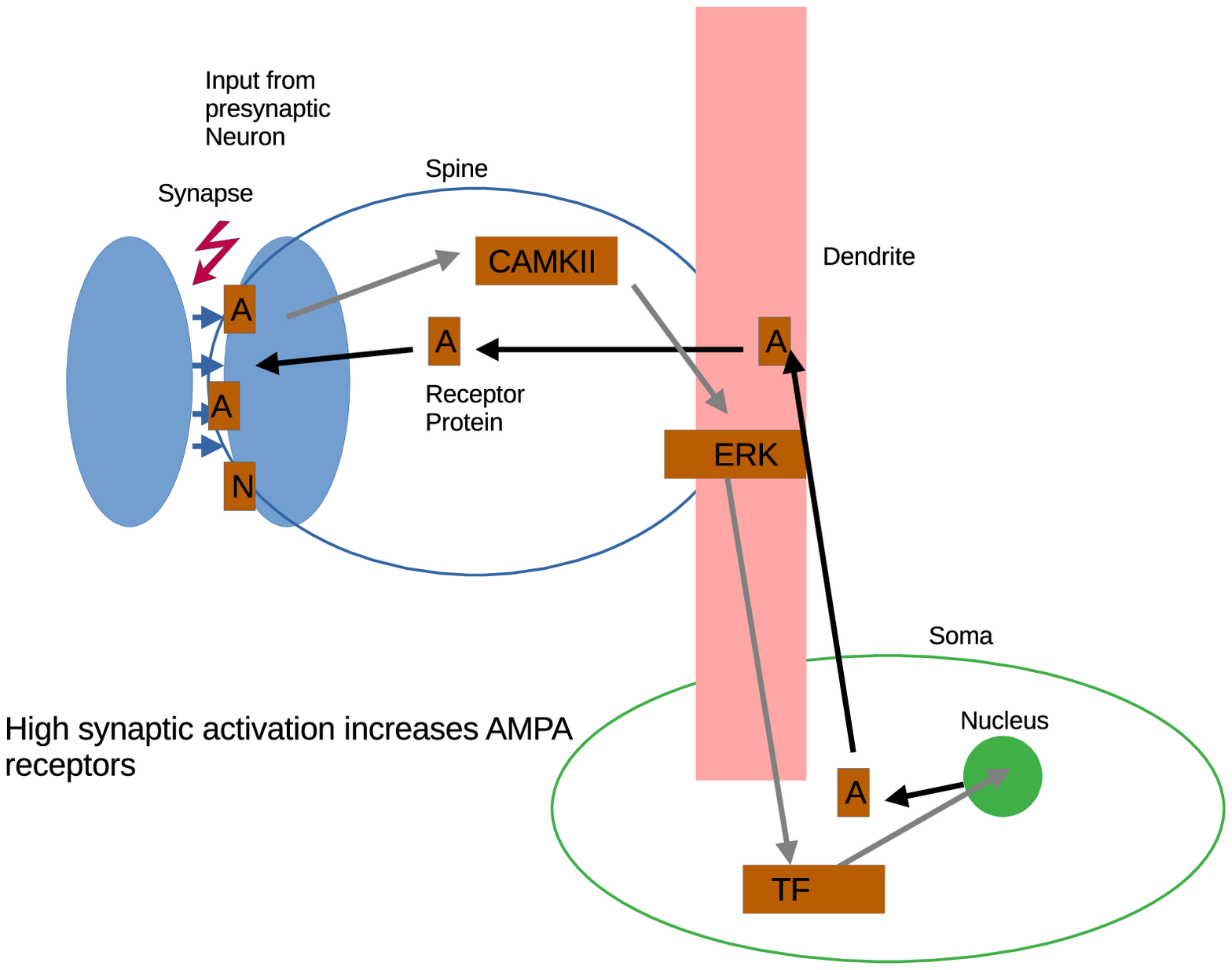}
\hspace*{-1mm}
{\bf B}
\includegraphics[width=0.45\textwidth]{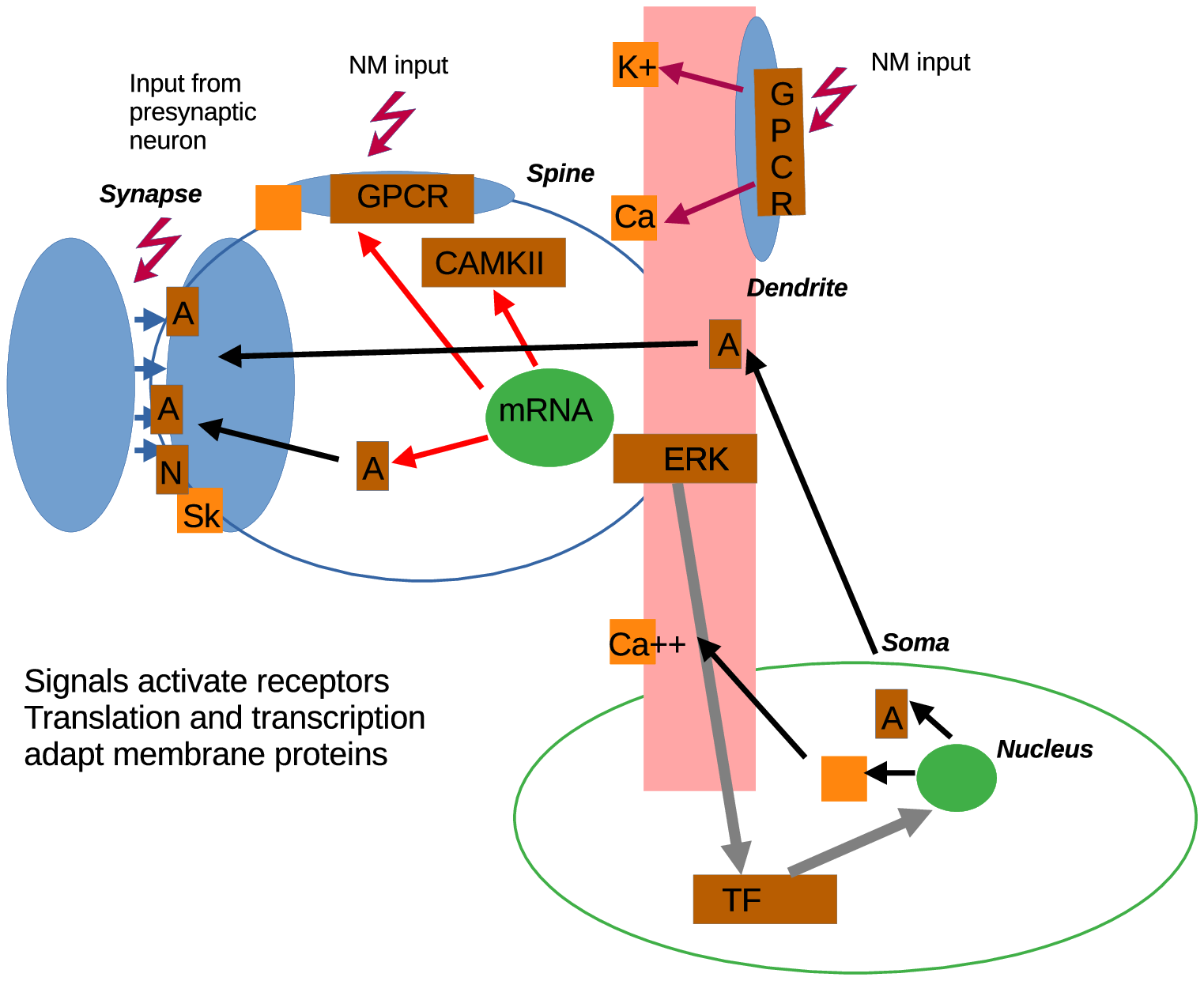}
\end{center}
\caption{({\bf A}) shows an outline of the classical synaptic plasticity model. AMPA [A] receptor and NMDA [N] in postsynaptic position receive signals from a paired, presynaptic neuron. Calcium influx will activate [cAMKII] and maybe other proteins, activating [ERK] and then transcription factors [TF] in the cytoplasm, which enter into the nucleus, where new AMPA [A] receptor protein is produced and transported back to the potentiated synapse.
({\bf B}) shows a more realistic picture, with input at the synapse from a paired neuron and input at synapse, spine, and dendrite from neuromodulation [NM]. Both mRNA in the spine and DNA in the nucleus can produce new proteins. GPCRs control ion channel activation, expression levels are DNA-regulated. Activated [ERK-TF] cause nuclear read-out. 
}
\label{fig:LTP}
\end{figure*}

The neuronal cell performs a number of cellular tasks and maintenance
mechanisms. Our goal must be to isolate information processing from
other complex cellular operations. Therefore, the systems of
sub-membrane and cytoplasmic intracellular signaling (\protect\hyperlink{ref24}{Bhalla and Iyengar 1999};
\protect\hyperlink{ref25}{Jarvis 2021}) may be replaced in a simplified
way by amorphous internal parameters, to be elaborated at a later time.

Experimentally, membrane receptors respond to signals from the outside,
but their localization and sensitivity is guided and controlled by
submembrane processes (\protect\hyperlink{ref20}{Scheler 2013},
\protect\hyperlink{ref26}{2004a}). This is true for ligand-gated ion
channels like AMPA/NMDA and GABA-A, as well as NM-gated G-protein
coupled receptors (GPCRs). We suggest to regard submembrane or
cytoplasmic proteins and small molecules (`second messengers') together
as internal parameters.

External parameters -- receptors and ion channels -- influence neural
transmission properties, which are visible to other neurons. Internal
parameters only have indirect effects on neural transmission and remain
hidden from other neurons. External and internal parameters together
form the `processing layer' of the neural plasticity system. In contrast
to this, the nucleus contains `core' parameters, both epigenetic
(histones) and genetic (DNA). This provides another layer of depth for
processing information, as long-term memory and possibly individualized
prior states (\protect\hyperlink{ref144}{Zhou and Walsh 2026}). Core parameters receive and send
information to the internal parameters via the nuclear membrane. In this
paper, we will not analyze core parameters and their function in detail,
we will focus on the actual processing layer instead. This is a position
paper which names principles and ideas that will have to be further
developed.

\hypertarget{principles-for-a-neuron-centric-model}{%
\subsubsection*{Principles for a neuron-centric
model}\label{principles-for-a-neuron-centric-model}}

External parameters roughly correspond to membrane proteins, such as
receptors and ion channels. They undergo plasticity on at least three
time-scales: fast (milliseconds, (de)sensitization), intermediate
(seconds-minutes, endocytosis, insertion), and slow (hours, new proteins
and morphological features like spines). Only the fast responses, which
consist of protein conformational changes, do not need the participation
of the internal processing system. The endocytosis/insertion of membrane
proteins requires a system of internal interactions. The long-term
ubiquitination of proteins and the generation of new proteins uses core
parameters.

The responses to signals by neuronal cells in the immediate domain has
been investigated as short-term (ST) synaptic plasticity. ST synaptic
plasticity is known to be highly differentiated, uses all 4 logically
possible types, is specific to neuron type and sometimes even synapse,
and also highly variable, sometimes changing response type even while
processing (\protect\hyperlink{ref27}{Zucker and Regehr 2002};
\protect\hyperlink{ref28}{Larsen and Sjöström 2015};
\protect\hyperlink{ref29}{McFarlan et al. 2022}). Presumably, there are
underlying mechanisms not apparent at the surface, which guide
responsivity and explain the variability, such as abundance and
availability of internal signaling proteins. The set of internal
parameters is the first line in orchestrating the plasticity of external
parameters. Both dendritic spines and axonal boutons contain subsets of
cellular proteins and RNA for local production and regulation, at least
about hundred protein species, which regulate membrane protein abundance
in the absence of signaling to the core (\protect\hyperlink{ref30}{Donlin-Asp et al. 2021};
\protect\hyperlink{ref31}{Kasai et al. 2021}). This allows limited
compartmental (not just synaptic), plasticity at the neurons which carry
spines, which are usually projection neurons with the strongest adult
plasticity.

Two examples of internal parameter activity are offered here:

\begin{enumerate}
\item
  \emph{calcium buffers}. Under high signaling activity, calcium
  influx into the cell is high, which activates a cellular braking
  mechanism via Sk/Bk-channels lowering the firing rate. However, with
  many available calcium buffers in the cell, the cell can absorb
  calcium influx for considerable time, store it with the available
  proteins and prolong high firing rates. Such effects have been used in
  neural models (\protect\hyperlink{ref32}{Rodrigues et al. 2023}),
  which used plasticity rules based on internal calcium-related
  parameters and regulation.
\item
  \emph{cAMP production}. The cAMP production system has fluctuating
  abundances of multiple variants of production enzymes (AC) and
  degrading enzymes (PDE). This means that GPCR activation during low
  cAMP availability will lead to a limited amount of downstream
  activity. Only very strong signals will have any effect. The concept
  of fluctuations without lasting plasticity applies to this domain.
  However, the neuron may undergo transforming events such as reading
  out new proteins, changes in morphology, and insertion of new receptor
  proteins at the synapse based on core activation, which translates
  into long-term memory.
\end{enumerate}

We hypothesize that internal parameters can be used to code for an
internal model of signal response. When neurons receive a variety of
signals at their membranes, they are not processed in a uniform way.
Rather, the processor itself is variable and dynamic state-dependent.
When a main signaling pathway is highly active, signals are transferred
to actuators -- such as ion channels -- and the system responds
externally, for instance by altering its firing rate, its responsivity
to fast signals, its receptivity to AMPA vs. NMDA etc. If a signaling
pathway is down-regulated, the same external signals have little effect.
This also affects plasticity. Highly active responses initiate new
protein expression via mRNA translation consistent with information
storage, while down-regulated pathways will block storage and prevent
memorization. This view is well justified from the available literature
(\protect\hyperlink{ref33}{Park et al. 2020}), but it is rarely taken
into account even when plasticity is explained as a set of molecular
events in the brain. A static view of a predictable, fixed chain of
events prevails (\protect\hyperlink{ref34}{Lisman et al. 2002};
\protect\hyperlink{ref35}{Wu et al. 2006};
\protect\hyperlink{ref36}{Hayashi and Huganir 2004}). In reality,
synaptic and membrane plasticity is not only and not directly dependent
on the signals between neurons. Instead, the activity of the internal
system is necessary for selecting and responding to signals (\protect\hyperlink{ref33}{Park et al. 2020};
\protect\hyperlink{ref37}{Alejandre-García et al. 2022}). The response
of the neuron cannot be determined exclusively from the signals it
receives. Its internal state matters.

\hypertarget{signaling-to-the-core}{%
\subsubsection*{Signaling to the core}\label{signaling-to-the-core}}

One task of the processing layer to induce lasting plasticity is to take
distributed, time-structured input signals and produce a spatially
centralized, temporally integrated signal to the core, i.e.
transcription factors which move to the nucleus and cause DNA read-out
of individual genes or genetic programs. This is not a simple task.

The internal system does not operate in an amorphous way. It contains
localized elements in distinct positions, at dendritic spines, at the
soma, at axonal boutons, at a synapse. It also has a central `workspace'
in the form of the cytoplasm with its organelles and protein complexes.
The long-term memory in protein expression which is caused by RNA
translation and especially DNA transcription could sometimes function as
a ``write-once'' event. Once it has been executed, the system retains
the new structure of protein abundances, essentially indefinitely. This
is permanence in long-term memory, an additional important
qualification. While new strong signaling may overwrite the structure,
such events may be kept very rare `permanent memory' by the internal
parameter settings.

Signals are selected at the membrane. They are then shaped and
re-structured in time. They need to be sorted or transferred to protein
complexes in the cytoplasm and ultimately activate the proteins which
enter into the nucleus. In the nucleus, epigenetic adaptation regulates
access to DNA, transforming signals by enhancement and suppression (\protect\hyperlink{ref38}{Josselyn and Tonegawa 2020}). It is clear then
that signals from the periphery have to be processed on several levels,
and with several types of control structures, before they can actually
be used to control or regulate DNA read-out, or be stored in the
epigenetic layer around it. In this way, whole genetic programs can be
started, such as morphological growth of spines, axons, or dendrites.
Individual genes (e.g. AMPA receptors, GPCRs) may also be transcribed
during periods of neural plasticity.

\hypertarget{signaling-within-the-processing-layer}{%
\subsubsection*{Signaling within the processing
layer}\label{signaling-within-the-processing-layer}}

We have discussed how the the internal parameter system sends signals to
the core for long-term and permanent plasticity. Its other task is to
enact short-term plasticity. Signals, fast glutamatergic/GABA synaptic
or slower GPCR activation, arrive at the membrane, where they are being
transformed through internal parameters by feedback cycles and similar
control structures (cf. \protect\hyperlink{f2}{Figure 2}). These can
suppress, lengthen, or augment a signal by engaging the intracellular
signaling network. A well-known example is concurrent activation of
calcium- and cAMP-pathways in some cases as a requirement for synaptic
weight increase (\protect\hyperlink{ref39}{Manninen et al. 2010};
\protect\hyperlink{ref116}{Scheler 2013}). The result is updates of external membrane
parameters resulting from a combination of internal parameters and
incoming signals (s. master equation Eq (1)).

\begin{figure*}[htb]
\noindent
\begin{center}
\includegraphics[width=0.8\textwidth]{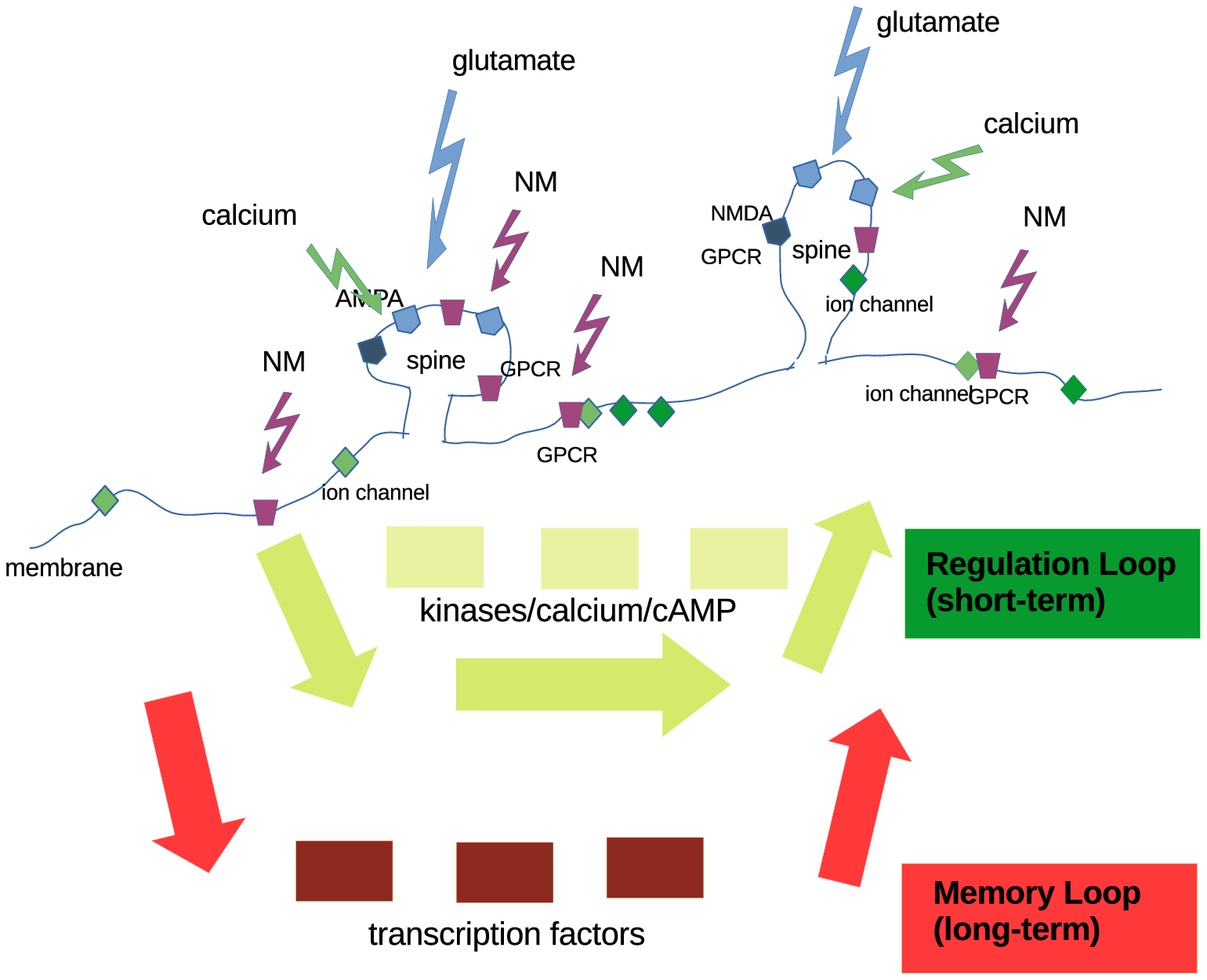}
\end{center}
\caption{Regulation and plasticity of membrane parameters. Ion channels and receptors at the spine or the extrasynaptic membrane receive signals and activate an internal system which (a) changes properties of receptors/ion channels short-term (regulation loop) or (b) continues to signal via transcription factors to the nucleus (memory loop). Activating the memory loop may require stronger signals and additional readiness on the part of the internal system.}
\label{fig:internal-filter}
\end{figure*}

The membrane with its many inserted proteins defines the neuron's spike
response, i.e. its activation function. This controls neural
transmission, i.e. the horizontal interactions between neurons. The
membrane is compartmentalized into dendritic branches, spines and
synapses, for incoming signals. Signals are therefore received by a
highly structured system. They can be located by the system, and
processed in a spatially distributed way. Synaptic glutamatergic
signals, which are brief (milliseconds), produce calcium transients
mostly from AMPA and NMDA receptors and voltage-gated calcium channels
(VGCCs). Calcium buffering proteins and calcium release from internal
stores (\protect\hyperlink{ref40}{Humeau and Lüthi 2007};
\protect\hyperlink{ref41}{Gulledge et al. 2005}) contribute to the
signal. Calcium is a major integrated signal for induction of plasticity
where different shapes of the signal result in different outcomes (\protect\hyperlink{ref42}{Ramaswamy and Markram 2015}). The selection
and filtering of membrane signals is performed by the internal system
with its parameters, e.g. availability of calcium-buffering proteins and
release from internal stores.

We assume that the internal system itself is a homeostatically closed
system (\protect\hyperlink{ref43}{Deng et al. 2011};
\protect\hyperlink{ref44}{Shinar and Feinberg 2010};
\protect\hyperlink{ref45}{von Bertalanffy 1969}). Plasticity occurs via
regulation of the abundances of protein species, via mRNA translation or
DNA read-out. In this way, the properties of the intracellular system
can be re-set and re-calculated. This observation strengthens the view
of the internal parameter system as a processing layer, programmable by
signals from the outside and updated from inside via parameter re-sets
such as local mRNA translation and the core DNA transcription system
(cf. \protect\hyperlink{f2}{Figure 2}).

To summarize, the membrane layer is adaptive on several time scales, and
has permanent features from epigenetics and genetic diversity. Under
specified conditions, protein abundances in the processing layer or at
the membrane change, adapting the neuron to a new environment. Long-term
potentiation/depression (LTP/LTD) can be seen as a special case of
signal-induced long-term membrane plasticity, after strong ongoing or
temporally structured stimulation, temporal integration for membrane
signals, and extended intracellular responses, such as engaging the
kinase/phosphatase system (\protect\hyperlink{ref46}{Hunter 1995}). In
general, the neuron regulates its responses, keeping itself intact as a
functioning cell, and adjusting to its signaling environment within a
network of other neurons.

\hypertarget{signal-selection-and-filtering}{%
\subsection*{Signal selection and
filtering}\label{signal-selection-and-filtering}}

\hypertarget{plasticity-of-membrane-parameters}{%
\subsubsection*{Plasticity of membrane
parameters}\label{plasticity-of-membrane-parameters}}

In our conceptual model, parameters stand for the expression levels of
proteins/molecules, sometimes for groups of them, or variants of the
same molecule. A precise matching of parameters to biochemistry is not
intended at this point. These parameters may sit externally, in synaptic
or non-synaptic positions at the membrane, internally near the membrane,
or localized somewhere in the cytosol, or they may be core parameters in
the nucleus. On a short time scale, external parameters mediate signal
response and internal parameters adapt the neuron's external parameters.
On a longer times scale, internal parameters are plastic. Neurons
receive and respond to signals which are first sensed by external
parameters. These occur mostly at synapses as the entry points for
signals. There are also receptors in non-synaptic dendritic positions,
where neuromodulatory (NM) receptors may act in a 'bulk' fashion
(``volume transmission'', (\protect\hyperlink{ref47}{Özçete et al.
2024})). Another variant is GABAergic input from chandelier cells (\protect\hyperlink{ref48}{Contreras 2004}) at the axon initial segment,
which act as gatekeepers as well as NM receptors controlling axonal
release. While synaptic signals are driving the spiking activity of the
neuron, they are also being processed with the help of internal
parameters, i.e. filtered or selected for further short-term computation
and long-term plasticity. Immediate and reversible plasticity of
receptors and ligand-gated ion channels (AMPA, NMDA, GABA-A) is the
first response. When receiving signals, these parameters adapt on a
short-term basis, e.g. by conformational changes in receptor proteins
(receptor sensitization and de-sensitization). In the absence of strong
signal selection, these adaptations are short-term reversible and the
effect may be described as fluctuation rather than adaptation. When
calcium influx and increase of small molecules, like cAMP, engage the
internal parameter system, in particular the kinase/phosphatase balance,
control structures and computations by internal parameters decide about
the fate of signals (cf. \protect\hyperlink{f2}{Figure 2}).

An important question is how signals are accumulated over space and
time. In many situations, external parameters will return to their
original values (`fluctuation'), but they may also retain new adaptive
values, when certain circumstances are met (`plasticity'). To achieve
long-term memory, new protein translation by mRNA read-out or
transcription from DNA is necessary. In \protect\hyperlink{f2}{Figure
2}, the slower loop, the 'memory'
loop engages transcription factors, which transmit to the nucleus. The
faster 'regulation' loop may occur within a dendritic spine, which
appears as a specialized substructure, almost an organelle for synaptic
adaptation. Dendritic spines (and possibly axonal boutons) act as
cellular compartments with a reduced but functioning intracellular
signaling apparatus (\protect\hyperlink{ref49}{Nimchinsky et al.
2002}).

There is a notable difference in plasticity between spiny and aspiny
neurons. Spines are localized at projection neurons in areas of strong
adult plasticity, such as pyramidal neurons in cortex, hippocampus and
amygdala, medium spiny neurons in striatum and in certain dendritic
regions of Purkinje cells in cerebellum -- areas of high fine-tuned
plasticity. For spiny neurons, spine removal and generation is an
efficient modification of existing information. In contrast, synapses of
aspiny neurons don't have a dedicated compartimentalized local system of
intracellular signaling. Aspiny plasticity relies either on immediate
local, or on global, rather than compartmental, intracellular parameters
by diffusion. In general, adult plasticity at aspiny neurons is less
articulated and less differentiated (\protect\hyperlink{ref50}{Renner
and Rasia-Filho 2023}).

To find out when neural transmission causes plasticity is not easy to do
in a general manner. Many conflicting data and effects have been
experimentally observed (see below). We hypothesize that membrane
proteins have a strong tendency towards homeostasis, i.e. return to
existing values, after short-term disruption (\protect\hyperlink{ref51}{O'Leary et al. 2014}). But in some situations,
the accumulation of traces from synaptic signaling events is dominant (\protect\hyperlink{ref52}{Abraham et al. 2024}). For the synaptic
ligand-gated ion channels AMPA and NMDA a number of experimental
protocols exist which cause lasting plasticity (up- or down-regulation,
(\protect\hyperlink{ref53}{Chen et al. 1999})). It is also known that
ion channels integrated in the membrane are variable in their expression
levels (\protect\hyperlink{f3}{Figure 3A,B}). The factors influencing
this, often summarily referred to as intrinsic plasticity (\protect\hyperlink{ref54}{Debanne 2009};
\protect\hyperlink{ref55}{Debanne et al. 2019}), are not well-explored
(cf. below). Calcium modulation has often been indicated as a major
factor in initiating synaptic plasticity (\protect\hyperlink{ref56}{Johenning et al. 2015};
\protect\hyperlink{ref57}{Nishiyama et al. 2000};
\protect\hyperlink{ref58}{Bloodgood and Sabatini 2007};
\protect\hyperlink{ref59}{Sabatini et al. 2002}), but there are many
other indications for necessary signal types and co-factors (\protect\hyperlink{ref60}{Dolmetsch 2003};
\protect\hyperlink{ref61}{Mellström et al. 2008};
\protect\hyperlink{ref62}{Fujii and Bito 2022}). G proteins are internal
proteins linked to neuromodulatory receptors (GPCRs (\protect\hyperlink{ref63}{Liu et al. 2024};
\protect\hyperlink{ref64}{Rockman and Lefkowitz 2024})), which have a
number of functions. They affect -- usually restrict -- transmitter
release at axonal boutons when placed in presynaptic position (\protect\hyperlink{ref65}{Scheler and Schumann 2003}), and transiently
regulate the ion channels in the membrane, reducing or enhancing their
efficacy. Ion channels determine the excitability of the neuron. NM-ion
channel interactions can reversibly remodel dendrites at intermediate
time frames (\protect\hyperlink{f3}{Figure 3B}, (\protect\hyperlink{ref66}{Furusawa and Emoto 2021};
\protect\hyperlink{ref67}{Kanamori et al. 2015})). This means neuronal
activation functions are conditional on which (central) NM signal
modulates them. The activation of different NMs highlights the set of
neurons that are modulated by them (\protect\hyperlink{ref65}{Scheler
and Schumann 2003}).

\begin{figure*}[htb]
\begin{center}
{\bf A}
\includegraphics[width=0.5\textwidth]{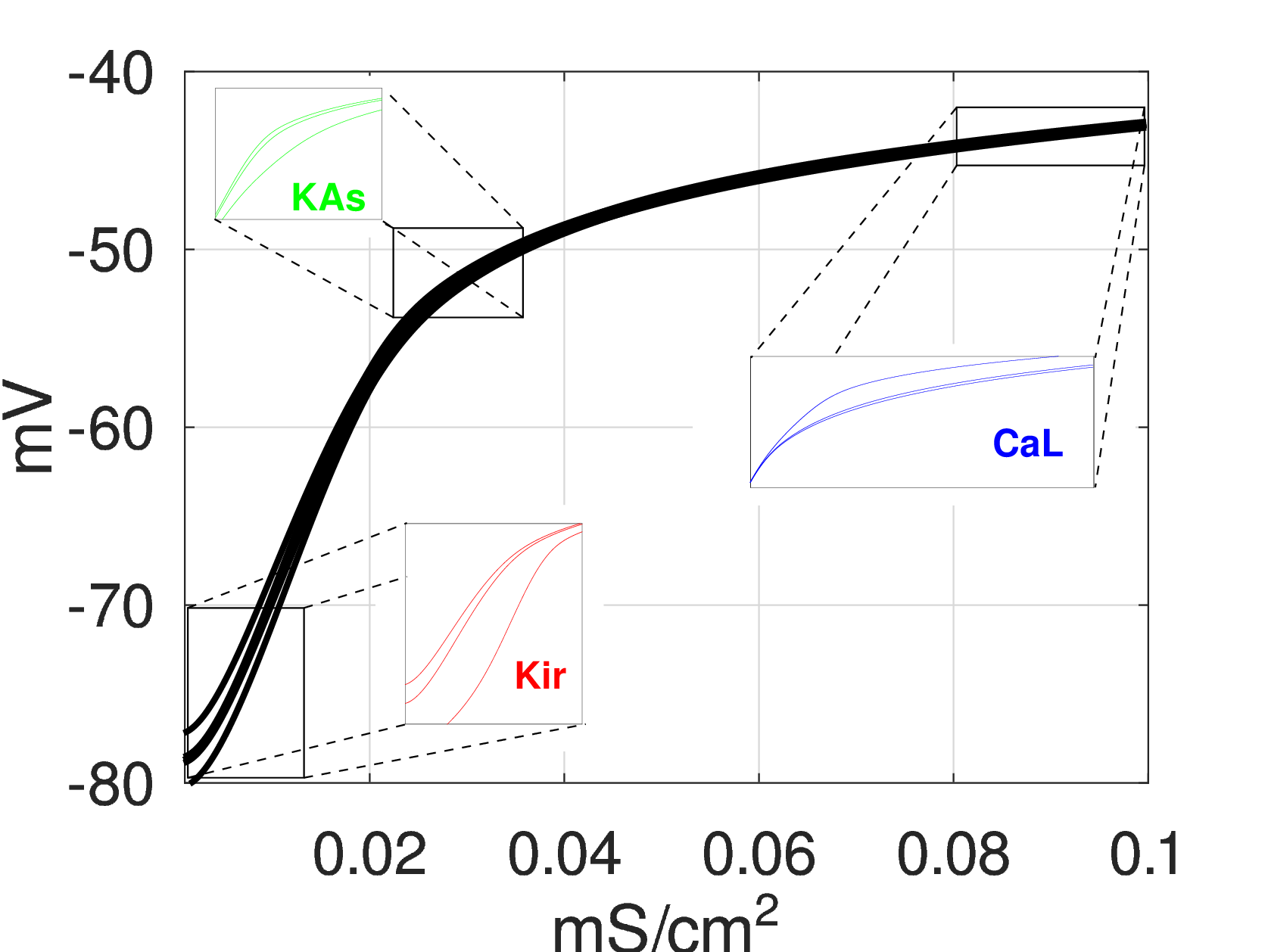}\hfill
\hspace*{-1mm}
{\bf B}
\includegraphics[width=0.40\textwidth]{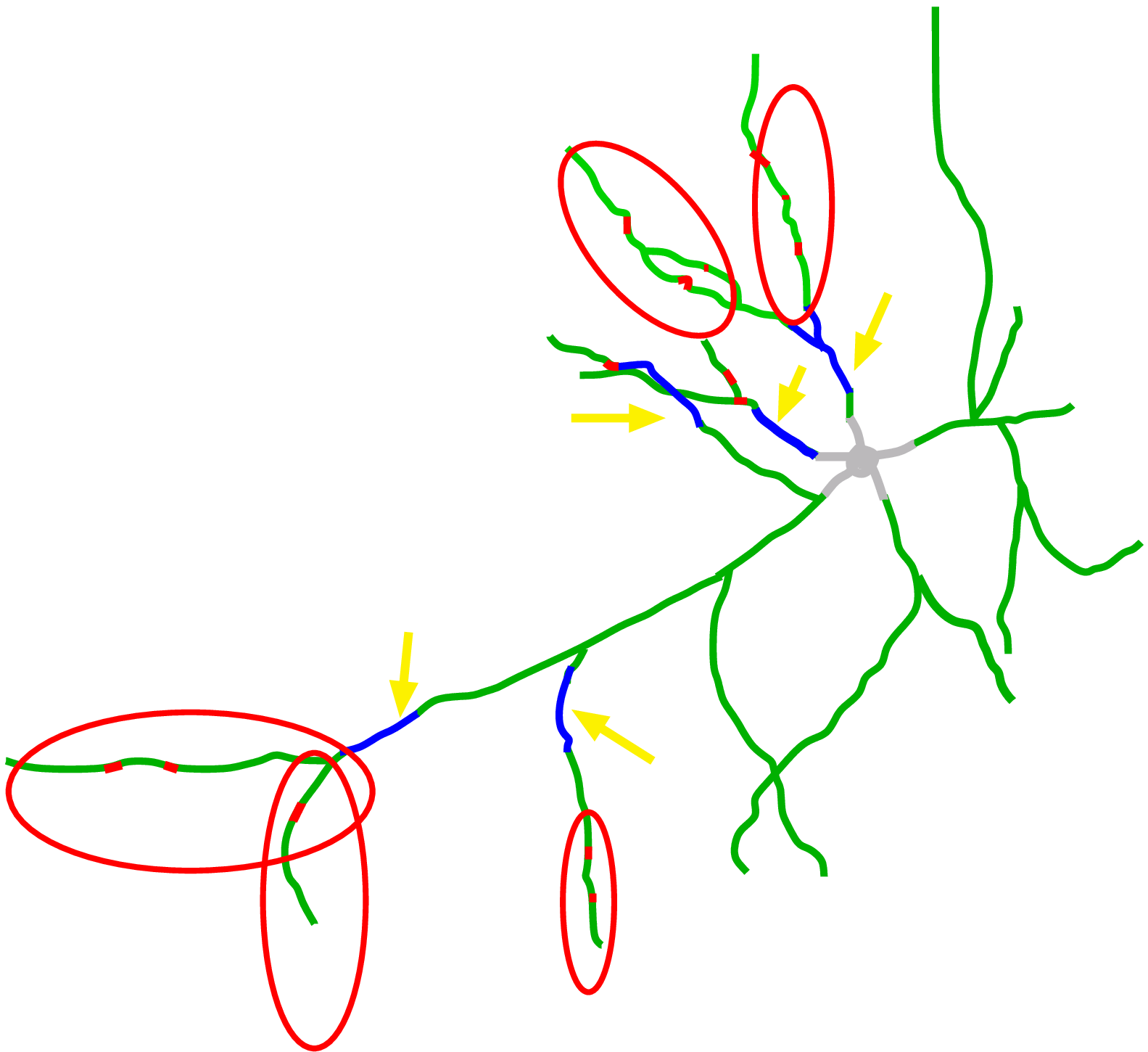}
\end{center}
\caption{{\bf A} shows the variability of ion channel density and the selective modulation of a neuron’s activation function. 3 different ion channels are analyzed, cf.  
(\protect\hyperlink{ref108}{Scheler 2014b}) for details.
{\bf B}  shows a model neuron with NM modulated re-design of its dendritic function. Red dots show putative spine inhibition (in red-circled areas), blue lines show the decreased dendritic transmission, marked by yellow arrows.
}
\label{figure-intrinsic}
\end{figure*}

\hypertarget{models-for-learning-rules}{%
\subsubsection*{Models for learning
rules}\label{models-for-learning-rules}}

It seems questionable that abstractions of synaptic and neural
plasticity could take on the general shape of a `learning rule'.

Many factors influencing outcomes have been experimentally observed:
neuromodulation, subsequent or concurrent; spacing of synaptic signals;
reinforcement by recurrence in a network (`re-entrant'); capacity of
internal neural plasticity pathways like CaMKII, PKA, or MAPK; amount or
strength of neuromodulation; the NMDA Mg++ switch, a non-linear effect;
concurrent GABA signaling; dendritic processing such as synaptic
positioning on dendritic branches; backpropagation of calcium signals to
distant synapses; dependence on ion channels gating backpropagation,
synaptic spacing (AHP); additional hyperpolarization (GIRK) or
depolarization (Musc, K~+~A- channels), cross-talk among internal
pathways; globality of the G protein pathway;and possibly others).
Intracellular calcium buffers and mechanisms for release from
intracellular stores influence calcium transients and are an example for
the integration of outside signals and internal parameters, with strong,
high calcium transients vs. lower ongoing signals. Sk-channels act as
plasticity blockers at a synapse. Neuromodulation (NM) via G-protein
coupled receptors (GPCRs) (\protect\hyperlink{ref68}{Shenoy and
Lefkowitz 2011}) also influences signal selection using presynaptic NM
receptors for gating.

However, it may be possible to postulate principles for a number of
learning rules which interact to produce an outcome in specific
well-described situations. Synaptic signals to a neuron have the
potential to activate transcription factors as immediate early genes
(e.g. Arc, CREB) within minutes (5--20~minutes for Arc in hippocampal
neurons (\protect\hyperlink{ref69}{Guzowski et al. 1999},
\protect\hyperlink{ref70}{2001})). These transfer to the nucleus (\protect\hyperlink{ref71}{Rienecker et al. 2022};
\protect\hyperlink{ref72}{Minatohara et al. 2016}), and start new
transcription. But they have to pass the intracellular signaling system.
This has a large number of control structures such as thresholds,
feedback loops, feed-forward loops, antagonistic signaling (\protect\hyperlink{ref105}{Scheler 2014c}). For instance, negative
feedback loops serve to broaden a signal and extend its duration. They
may also dampen a signal below threshold (\protect\hyperlink{ref74}{Scheler 2012}). Signals may be modified, such
that later signals in a sequence are being suppressed. In contrast, a
sequence of signals may be enhanced to reach a threshold (\protect\hyperlink{ref75}{Maus et al. 2020}). These control structures
are activated by signals but they depend on internal parameters. Control
structures may use existing neighborhood structures, such as the
neighborhood relations on the dendrite, to regulate signals (\protect\hyperlink{ref76}{Gidon et al. 2020};
\protect\hyperlink{ref77}{Winnubst and Lohmann 2012};
\protect\hyperlink{ref78}{Makino and Malinow 2011}), either potentiating
them, or canceling them out against each other. In this sense, internal
control structures serve to adapt a signal and influence whether it is
passed on to the next layer.

The synaptic model of plasticity assumes that an activated synapse
signals to the core, and the core signals back to the relevant synapse.
This ``synaptic tag'' hypothesis has been much discussed (\protect\hyperlink{ref80}{Rogerson et al. 2014};
\protect\hyperlink{ref81}{Redondo and Morris 2011};
\protect\hyperlink{ref82}{Li et al. 2014};
\protect\hyperlink{ref924}{ MartinK1997}). The
main problem it attempts to solve is how to separate synapses in
different positions, when they all signal through similar pathways 
(\protect\hyperlink{f1}{Figure 1}). It is difficult not only to maintain spatial
separation but also temporal distinction since internal processes
integrate signals over time. A molecular ``tag'' that would solve the
issues of maintaining strict synaptic separability has not been
identified.

In contrast, we may assume that the intracellular network serves to
integrate signals in space and time, and not maintain synaptic position
under all circumstances. This would speak against a pure ``Hebbian''
synaptic update as a mechanism of information storage (s.
\protect\hyperlink{ref109}{Scheler 2017}) for Hebbian plasticity as a homeostatic adjustment
mechanism), but make synapses more dependent on the available local
protein signals.

Under conditions to be specified, local synaptic signals are translated
into signals to the intracellular network. The number of activated
synapses should be low at any time under a strong selection of
transmission events which leave a trace. and update events are centrally
regulated in an integrated way and would be essentially sparse. A strong
selection of signals before plasticity is initiated is highly likely
because adaptation is metabolically costly (\protect\hyperlink{ref145}{Dewitt et al. 1998};
\protect\hyperlink{ref146}{Inoue et al. 2026};
\protect\hyperlink{ref147}{Xu et al. 2026} 
).

Criteria for signal selection could be strength (high amplitude,
co-incident at various sites), repetition (at the same site), or long
duration (even at a lower level of signaling). The responsibility of the
intracellular (calcium, cAMP, etc.) system would decide the outcome.
Even fairly weak signals could be memorized, if signals combine within a
specific time frame, and if the neuron is highly responsive. Strong
signals may fail to register, if the signal is isolated and the neuron
has low responsivity.

Several models are possible:
\begin{enumerate}
\def\labelenumi{\arabic{enumi}.}
\item
  We may assume that external parameters fluctuate in response to
  signals, they follow a random walk and need reinforcement by internal
  parameters to store a value. Such a model could emulate modern feature
  selection methods (\protect\hyperlink{ref83}{Guyon and Elisseeff
  2003}) very effectively.
\item
  External signals could be selected by their magnitude and repetition.
  For instance, an individual spike may not cause plasticity, but a
  number of neuronal spikes which occur consecutively may cause
  plasticity, also if the pattern of such spiking recurs with breaks
  over time (\protect\hyperlink{ref84}{Wu and Maass 2025}). This would
  be suitable to store events which occur within short time frames and
  are being replayed internally. In this way a sensation or an
  internally generated pattern could be stored by recurrence.
\item
  External signals are stored if the internal system is highly
  responsive with main plasticity pathways (e.g. kinases) all ready to
  be turned on. After a neuronal or synaptic plasticity event, the
  system turns unresponsive (e.g. high phosphatase activation) and for
  periods of time the probability of further storage is low and may not
  even return to full naive responsivity. Experimentally, the effect of
  responsivity has been studied along these lines by (\protect\hyperlink{ref85}{Zhou et al. 2009};
  \protect\hyperlink{ref33}{Park et al. 2020};
  \protect\hyperlink{ref86}{Dehorter et al. 2015}).
\end{enumerate}

External signals could be integrated in 2. on faster time scales and
suppressed for slower time scales. But how processes are combined is not
decided. The common theme is that signals are ignored or passed on
depending on a combination of the responsivity of the neuron and the
strength of the signal. In spiny projection neurons, the spine may act
independently of the main cell body within the regulation loop.

\hypertarget{internal-computation}{%
\subsection*{Internal computation}\label{internal-computation}}

\hypertarget{neuronal-heterogeneity-and-adaptation}{%
\subsubsection*{Neuronal heterogeneity and
adaptation}\label{neuronal-heterogeneity-and-adaptation}}

Within the paradigm of a neuron-centric view of plasticity, each neuron
represents a processor which will process input information according to
its internal functions and current state. It is important to realize
that neural computations are performed by heterogeneous neurons with
different genotypes and cellular identity. They may therefore inherently
be more or less receptive to plasticity specific for different
parameters (e.g. AMPA vs. GPCR).

Recent work on neuronal genotypes has often confirmed that neuronal
connectivity is highly correlated to its genetic make-up (\protect\hyperlink{ref87}{Planert et al. 2025};
\protect\hyperlink{ref144}{Zhou and Walsh 2026}). This
emphasizes a significant restriction on adult plasticity, since a large
part of connectivity is not set up by data or patterns, instead it is
genetically determined. Perceptual input may be necessary to trigger
waves of activity to establish the synaptic connectivity in each case.

Evolution has done a major part of the work in establishing the neuronal
types. For instance, there are cortical, hippocampal, amygdalar and
other kinds of pyramidal neurons. Within cortical neurons, there are
dozens of genotypes which can be clearly distinguished (Chang and Luebke
2007; \protect\hyperlink{ref89}{Kim et al. 2015};
\protect\hyperlink{ref90}{Mao and Staiger 2024};
\protect\hyperlink{ref91}{BRAIN Initiative Cell Census Network (BICCN)
2021}; \protect\hyperlink{ref92}{Gouwens et al. 2019}). To model
neuronal types and their connectivity, the evolutionary processes do not
need to be re-created. Instead, we may set up neuronal types and regard
them as type-specific programmable units. This should greatly reduce
space-time complexity for adult learning. Nonetheless, our understanding
of the existing genotyped structures for cortical (or hippocampal,
striatal, etc.) functions is still developing.

We hypothesize that neurons have generic programs for signal induced
plasticity. This is modified by the existing epigentic and genetic core
parameters and matched to the maturity of the neuron. The complex
control structure for protein signaling may undergo adaptation to
acquire a specific internal type.

When undergoing plasticity, neurons may acquire an adapted program (\protect\hyperlink{ref37}{Alejandre-García et al. 2022};
\protect\hyperlink{ref93}{Paik et al. 2020};
\protect\hyperlink{ref94}{Traunmüller et al. 2025}) which then becomes
part of their activation properties. There is an indication for this in
\protect\hyperlink{ref95}{Bulovaite et al. (2022)}, where the lifetime
distribution of PSD-95, a postsynaptic protein, was tracked on neurons
in cortex and hippocampus. Young or innate (`naive') neurons and aged
(`mature') neurons have different lifetimes for PSD-95. This points to a
loss of flexibility in synaptic learning during maturation and aging, or
vice versa, an increased stability for learned information.

Default or generic plasticity configurations possibly contain random
elements. Exposure to patterned signals lets them acquire a specific
dynamic state which reflects the neuron's experience and expectation.
This is a 'mature' neuron, or 'adapted' neuron, which we can probe for
its behavioral response pattern (\protect\hyperlink{ref96}{Quiroga
2012}; \protect\hyperlink{ref97}{Sharma et al. 2024};
\protect\hyperlink{ref98}{Waidmann et al. 2022}). A mature neuron may
still re-learn, and possibly even re-set to the naive state under
exceptional circumstances. In the naive neuron there is no
individualized control structure to filter the signal. A neuron may
contain both naive and mature sites (especially at spines), it could be
`chimeric' with respect to maturity. A naive site will store or read in
a value, a mature site already contains a value, compares a new value
with the existing one, and it can be protected against overwrite
(Sk-channels). Patterned signals have the effect of imprinting a neuron
with a specific local control structure which keeps the state of the
neuron and its synapses in a certain configuration. Many behavioral
learning events may fulfill the criteria of such an imprinting effect on
naive neurons - there is also evidence that the brain, or at least the
cortex, keeps many naive neurons available even in adults for new
learning of situations and events (\protect\hyperlink{ref148}{Ovsepian 2019}). Imprinting
or instantiation of a neuron involves setting-up an internal parameter
set as a reflection of the external parameters. This internal set would
be a generic or generalized version of membrane parameters and could be
used to gate or filter external parameter changes.

It is important to realize that for many cognitive operations the unit
of processing will be neither the individual neuron nor its spines, but
groups of neurons which together form a concept, or unit of computation
at a higher level (see Section~\protect\hyperlink{ref24}{"Selection of ensembles and plasticity"}). The
individual properties of the neurons help to explain how these larger
units can develop and function.

\hypertarget{programming-a-neuron}{%
\subsubsection*{Programming a neuron}\label{programming-a-neuron}}

We showed that neuronal cells have internal signaling mechanisms which
are activated by external signals. The main known pathways for neural
plasticity are the calcium-related, the cAMP-related, and the MAPK/ERK
pathways. Other pathways, such as BDNF-Trk interact with them or signal
through other means (such as NF-KB (\protect\hyperlink{ref101}{Schirò
et al. 2022})). The interesting observation is that these pathways
fulfill the idea of a programmable system. The system consists of
interactions with characteristic time-scales and strengths, expressed by
kinetic parameters. By activating/de-activating proteins (e.g. by
kinases/phosphatases) the system is plastic on the seconds-to-minutes
time scales. By changing the abundance of protein species there is
long-term plasticity in the system. Connections may be strengthened or
disappear beyond a lower threshold based on the available numbers of
protein molecules.

An example for this is 'transactivation', or an \emph{overflow switch},
as a control motif in computation cf. \protect\hyperlink{f4}{Figure 4A}.
In \protect\hyperlink{ref102}{Scheler (2006)}, we showed how high
protein expression of a target pathway allows a receptor to activate a
target protein, which is not involved in signaling by this receptor at
baseline. This pathway is activated conditionally, i.e. only when
internal conditions are met, and when the signal at the receptor is
sufficiently strong. This is a clear case where signal + internal state
interact to determine the outcome. A suitable model could multiply
signal and protein expression linearly, or it could contain a non-linear
element, such as a threshold. This is what we assumed in
\protect\hyperlink{ref102}{Scheler (2006)}, when we talk of a switch
that becomes activated.

\begin{figure*}[htb]
\noindent
{\bf A}
\begin{minipage}[b]{0.45\textwidth}
\includegraphics[width=1.2\textwidth]{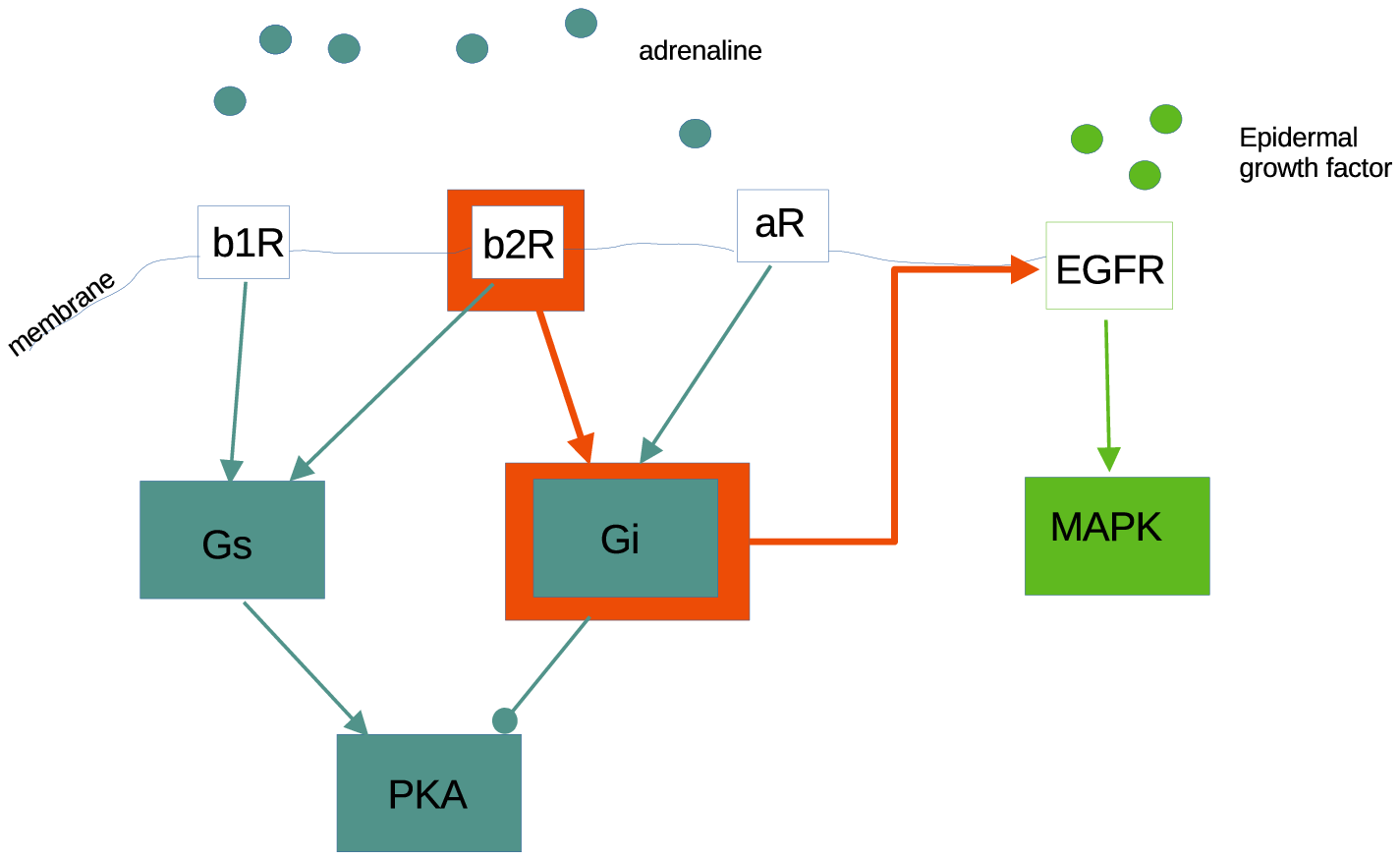}
\end{minipage}
{\bf B}
\begin{minipage}[b]{0.51\textwidth}
\includegraphics[width=\textwidth]{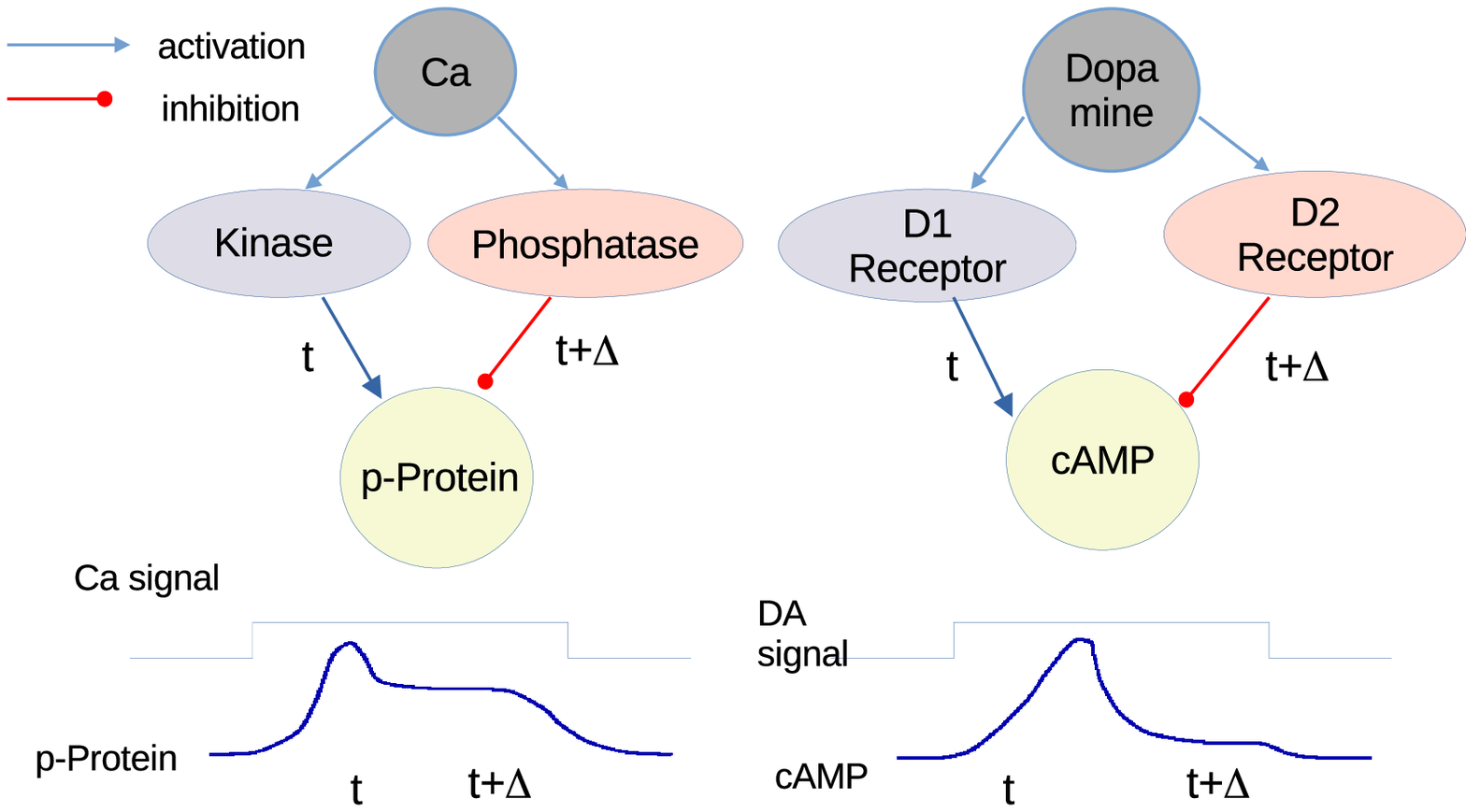}
\end{minipage}
\caption{Control Structures in Internal Computation.{\bf A}: Overflow switches 
(\protect\hyperlink{ref106}{Scheler 2006}).
Reorganizing structure in intracellular signaling. The orange pathways are only activated with high expression and activation of the beta-adrenergic 2 receptor and the G-protein Gi. Signaling via MAPK pathway from adrenergic activation at b1,b2 and a1 receptors can be switched on or off via the orange pathway. {\bf B}: Antagonistic signaling 
(\protect\hyperlink{ref105}{Scheler 2014c}).
A signal (calcium, dopamine) has both positive and negative effects on a target, often with a time difference. This results in fine-tuning of transient peaks at $t$ and the ongoing signal at ($t+\Delta$).}
\label{fig:control-structures1}
\end{figure*}

Another common motif for intracellular computation is \emph{antagonistic
signaling}. There are many examples for this. For receptors such as
D1/D2-type dopaminergic signaling, or beta/alpha-adrenergic signaling,
the signal input is linked to its output by two antagonistic pathways
(positive and negative) (\protect\hyperlink{ref105}{Scheler 2014c}).
The antagonistic motif allows re-scaling of inputs which vary over a
large dynamic range, compressing multiple orders of magnitude into a
single-scale response. Also, antagonistic signaling can generate peak
transients when one pathway is slightly faster than the other (\protect\hyperlink{f4}{Figure 4B}). We can modify the transient peak, or
the amplitude of the outcome of both pathways by protein abundance, i.e.
long-term plasticity.

Internal signaling can also reset presynaptic vesicle pools (e.g. by
cAMP-dependence), and it controls spine maturation and spine decay. It
performs spatial integration on the spine (e.g. when AMPA receptors
reservoirs on the spine shaft move to the spine synapse (\protect\hyperlink{ref103}{Araki et al. 2010})). There are local
membrane-near calcium buffers (e.g. CaMKII), which can store synaptic
signals for minutes by autophosphorylation of the kinase in the synaptic
positions where they are deployed. This leads to timed delays, queues,
and feedback cycles, which suppress a signal or enhance it and the
position and significance of which can be learned.

A significant specific property of internal signaling is the 'shared
workspace' of the cytoplasm which provides an access system for
dendritic and somatic compartments. Inputs from postsynaptic signaling
are sometimes routed into a shared workspace, from where they can access
the nucleus and induce new DNA read-out. At spinal synapses, internal
signaling may remain localized (\protect\hyperlink{ref104}{Steward and
Schuman 2003}) or spread within minutes along the dendrite. There is
also spatial integration over several dendritic compartments (\protect\hyperlink{ref68}{Shenoy and Lefkowitz 2011}) which can cause
activation or de-activation of a dendritic branch. The 'workspace' is
not homogeneous, but subspaces have distinct probabilities for molecules
to interact. Signals may being fed back to nearby locations , such as
closing ion channels on specific dendritic branches. Alternatively
signals may remain local. However, when signals are received at multiple
locations, there is then signaling via the shared workspace, and a
system-wide response.

An example for the power of the `shared workspace' is shown in
\protect\hyperlink{f5}{Figure 5A}. The system is versatile enough with
its many control structures to perform interesting functions. Using
different types of buffers available, we can even achieve a re-sorting
of signals in time. For instance, CaMKII which buffers calcium by
autocatalytic activity, releases calcium depending on the activation
status of each buffer molecule. This allows to create a reverse sequence
of calcium transients, which may be supported by internal calcium
release patterns. Individual synapse/spine input order may be reversed
according to internal properties, providing a basis for queued access to
nuclear transcription. A sequence of synaptic inputs, which are sorted
and combined from their spatial distribution into larger components
(e.g. like a bit code which is translated into `words'), could provide a
`canonical' shape of signal that can be easier read.

\begin{figure*}[htb]
\noindent
{\bf A}
\begin{minipage}[t]{0.40\textwidth}
\includegraphics[width=1.2\textwidth]{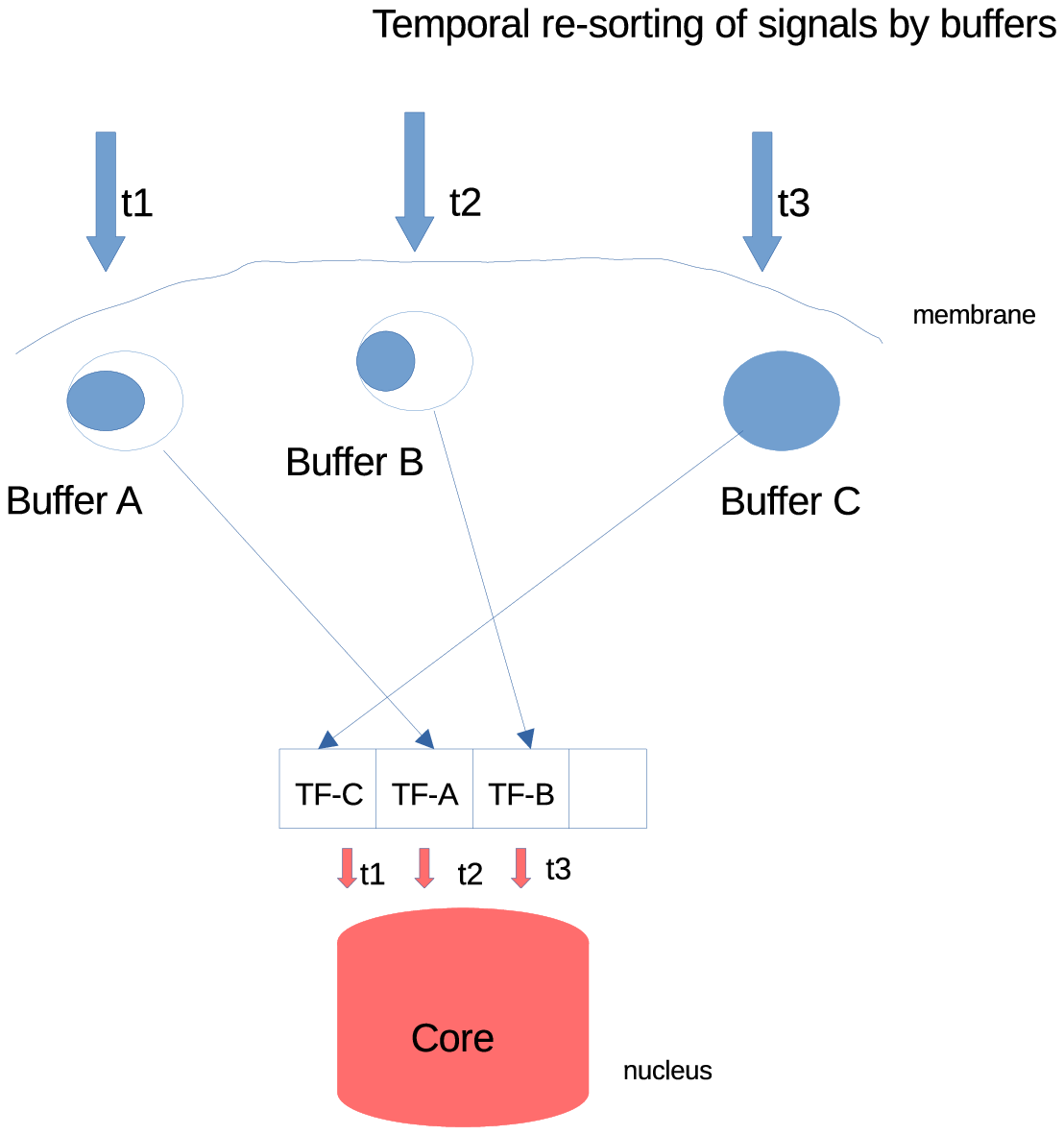}
\end{minipage}
\hfill
\ \hspace*{-12mm}
{\bf
 B}
\begin{minipage}[t]{0.66\textwidth}

\ \ \ (B1)\includegraphics[width=0.40\textwidth]{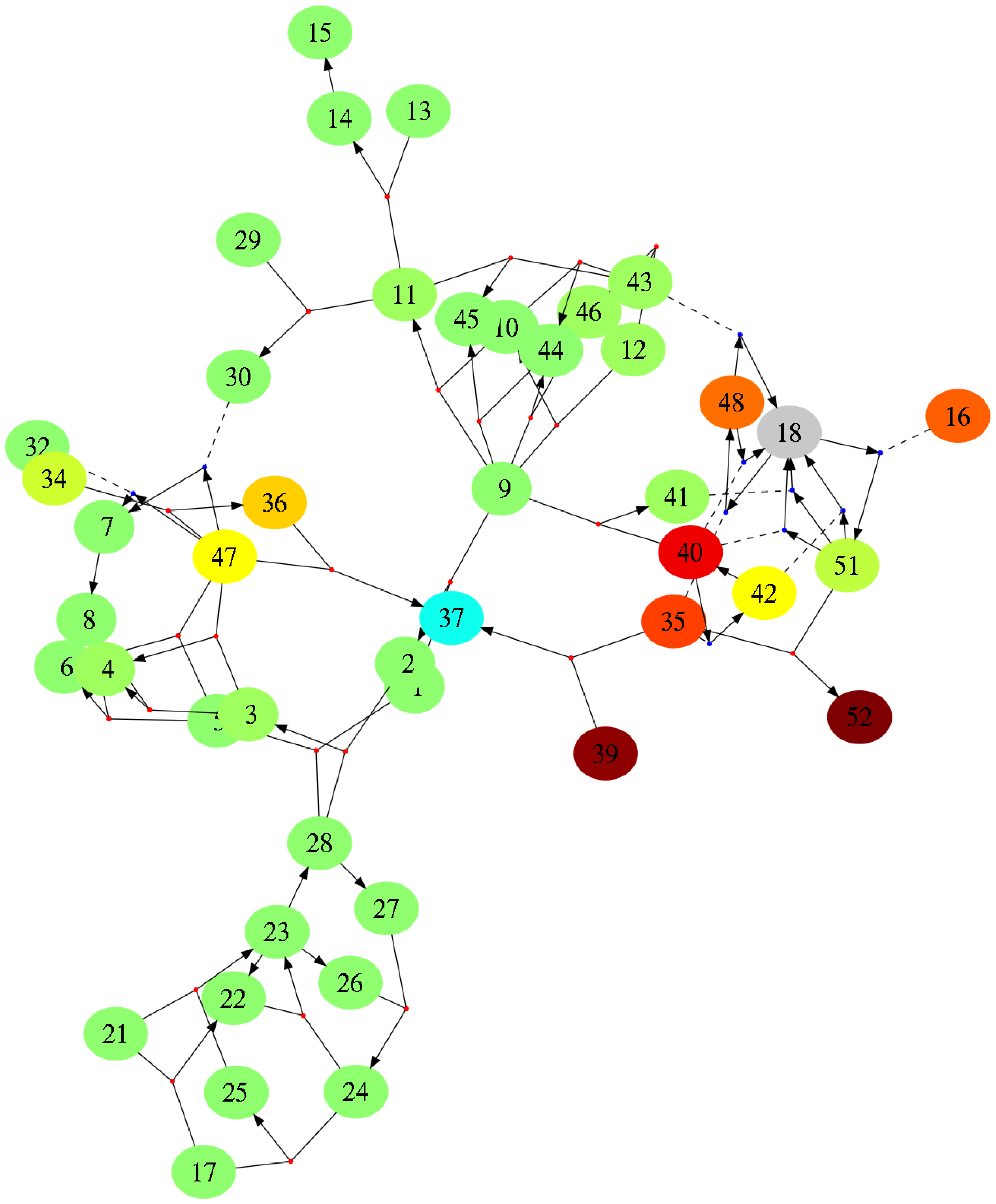}\ \ (B2)\includegraphics[width=0.40\textwidth]{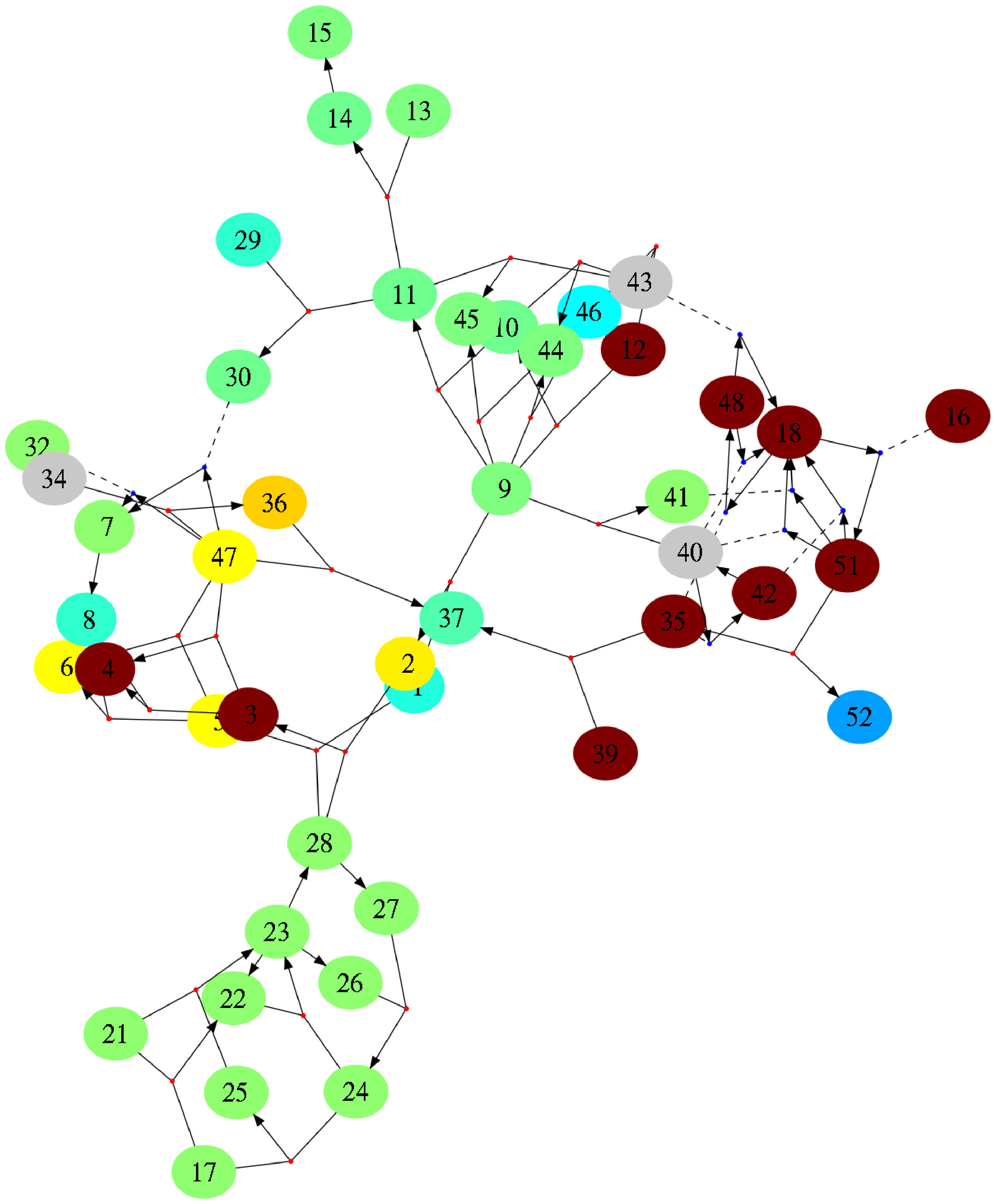}
\end{minipage}
\caption{Timing in Internal Computation. {\bf A}: Temporal re-sorting shows how buffers can re-order the timing of signals for activating transcription factors (TFs). Buffer C reacts first because it is saturated. {\bf B}: Changing the concentration of proteins in a biochemical reaction network (by 100\%, details in  
(\protect\hyperlink{ref128}{Scheler 2014b}).
Speed changes are shown in red (faster) and blue (slower) hues. (B1) Increasing a target protein (DARPP-34) has a narrow clustered response. (B2) Increasing both kinase (PKA) and phosphatase (PP2A) leads to a strong speedup of a larger number of reactions.
}
\label{fig:control-structures2}
\end{figure*}

For long-term plasticity, protein levels establish (by high expression)
or remove (by low expression) kinetic interaction connections between
pathways via their kinetic interactions. They also determine the
\emph{temporal execution} of the biochemical reactions underlying
intracellular signaling (\protect\hyperlink{ref79}{Scheler 2014b}).
High expression means that reactions are fast and homeostatic balances
are reached in short time. When protein expression is low, internal
reactions are slowed down, internal connections cut, and many signals
fail to be transmitted.

To summarize, signal input is received by external parameters at the
membrane and activate a program in the internal system which determines
the neural plasticity response (\protect\hyperlink{ref106}{Scheler
2005}). The internal computational system that filters or temporally
re-sorts signals, that combines signals, and gates them, is a
programmable and re-programmable system. The system contains parameters
which program its functionality which are being set by the core system.
To re-program the system requires new transcription or mRNA translation.
It can be reset by the core, a genetic read-out system. In (\protect\hyperlink{ref20}{Scheler2013}), we employed a
\emph{transfer function} approximation to generate look-up tables for
endpoints of internal parameters in response to outside signals. On this
basis, internal parameters can be acquired and adapted by setting
outcomes. Such a formulation may provide the basis for a learnable
system.

\hypertarget{selection-of-ensembles-and-plasticity}{%
\subsubsection*{Selection of ensembles and
plasticity}\label{selection-of-ensembles-and-plasticity}}

The formation of neuronal groups (ensembles) is prominent in various
areas of the brain, e.g. in mammalian cortex, where neurons are arranged
in layers and microcolumns (\protect\hyperlink{ref149}{Herculano-Houzel et al. 2008};
\protect\hyperlink{ref150}{Douglas et al. 2004};
\protect\hyperlink{ref151}{Horton et al. 2005})
or in striatum (\protect\hyperlink{ref}{Brimblecombe and Cragg 2017})
where there are patches.
These ensembles are usually comprised of projection
neurons interspersed with inhibitory interneurons, no matter whether the
projection neurons are excitatory or inhibitory. In many neural areas,
e.g. hippocampus, cortical regsions, we see widespread
electrophysiological activation of 30--50\% of neurons within a relevant
brain area in response to a significant behaviorally event, like a fear
experience. We also often see activation of immediate early genes (IEGs,
e.g. Arc) at a similar density, however, transient activation of IEGs is
not a reliable indicator of lasting neural plasticity (\protect\hyperlink{ref107}{Robbins et al. 2008};
\protect\hyperlink{ref108}{Bruins Slot et al. 2009}). Everyday
memorization of facts and data may rely on a much lower proportion of
neurons being activated, and use highly restricted localized plasticity
(\protect\hyperlink{ref109}{Scheler et al. 2025}).

Especially in cortex, input patterns face the task of finding the best
match among the existing structure. There are experimental data (\protect\hyperlink{ref110}{Sherman and Usrey 2024a},
\protect\hyperlink{ref111}{2024b}; \protect\hyperlink{ref112}{Usrey and
Sherman 2019}) that thalamic input activates many cortical ensembles in
parallel and selects the strongest matches by activations which are
recurrently potentiated, and probably connected via synchronization (\protect\hyperlink{ref113}{Scheler 2018};
\protect\hyperlink{ref153}{Fasoli et al. 2018}), while weaker matches are
suppressed and fail to become activated. A network of activated cortical
ensembles gives rise to representational functional states which are
measurable by brain imaging.

Such input patterns undergo a transformation via the signals they
produce at individual neurons and their integration into an ensemble. In
this sense, a whole ensemble is selected for memorization of a pattern,
not just a single neuron (\protect\hyperlink{ref38}{Josselyn and
Tonegawa 2020}). When a pattern is matched to an existing stored
template, it becomes adapted to existing knowledge. Strong signals may
generate internal plasticity and re-program a neuron. Otherwise,
internal inaction will stop the outside signal from passing into the
plasticity-inducing processing layer. The plasticity response of an
ensemble is therefore expected to be heterogeneous with respect to Its
neurons.

How can the plasticity of the single neuron be coordinated over a whole
ensemble? In an earlier paper (\protect\hyperlink{ref113}{Scheler
2018}), we provided a model, where neurons function either in a
synchronized transmission mode or in an irregular-asynchronous mode,
where they can read out and adapt membrane parameters. Input pattern
selection would result in high activation of an ensemble and
synchronization across ensembles. Such a regime could be described as
high 'strain'. (`Strain' could be defined and measured by both
synchronization and amplitude of a neuronal group.) This would
correspond to a high value for internalization. Parameter adjustment
(internalization) would occur in such an environment. It would cause a
network to internalize information from the external membrane, and to
read out from internal parameters so as to reduce the high strain state.
This would allow a return to a well-adjusted regime of low strain over
an ensemble. In this way the neurons within the ensemble could
automatically adjust their parameters to match the ensemble's task.

The interesting thing is that we may use the concept of strain to solve
the question of when to retain information and when to discard it for a
whole ensemble. When amplitude and synchronization (=strain) are high,
information is retained and features are relegated to each neuron. Under
low or relaxed strain neurons enter into a recurrent mode and adaptation
is read out. The ensemblelearning goal or the objective function can be
defined for the ensemble.

Neuromodulation has an specific role in that it codes for neuronal
identity in a combinatorial fashion (\protect\hyperlink{ref26}{Scheler
2004a}). Each neuronal type is differentiated by the NM receptors it
carries, and neurons may even be unique on an individual level in terms
of the localization of the NM receptors (\protect\hyperlink{ref154}{Scheler 2026}).
NM receptors can thus label both neurons and connections (nodes and
edges of a graph structure). The label states what NM receptor, if any,
is present (graph 'coloring'). Such labeling is extremely useful because
it allows for constrained inferences (\protect\hyperlink{ref65}{Scheler
and Schumann 2003}). For instance, all neurons labeled for dopamine D2
receptors hyperpolarize (via GIRK channels) at a dopamine signal, while
neurons labeled with D1 abolish afterhyperpolarization (AHP) and fire at
higher rates. This dichotomy is well-known and described for striatal
neurons (\protect\hyperlink{ref155}{Gerfen 2023}), along with its function.

Control structures using internal parameters in the membrane-near
intracellular layer influence external membrane properties directly and
immediately. Spatial resolution is kept during the fast adaptation of
neural membrane properties. Additionally, membrane-near internal control
structures select strong signals for plasticity. These are internalized
and maintained (e.g. with the help of calcium buffers, widespread
protein phosphorylation, increase of cAMP, inositol etc.) When the
signal reaches cytoplasmic protein complexes and transfers into the
nucleus, spatial resolution is lost and global neural remodeling occurs.
This is long-term plasticity and it builds on the existence and
diversity of core parameters.

A network of processors with their own individual memory, which are
connected by graded signals and which participate in low-dimensional
representational configurations can be expected to offer a highly
flexible and sophisticated framework suitable for complex computations (\protect\hyperlink{ref12}{Gallistel and Matzel 2013};
\protect\hyperlink{ref9}{Langille and Gallistel 2020}). Not only can we
compress weighted graphs considerably, 
we can
also expect to perform additional cognitive tasks automatically (e.g.
learning a category from a single example).

\hypertarget{mechanical-transmission-without-plasticity}{%
\subsubsection*{Mechanical transmission without
plasticity}\label{mechanical-transmission-without-plasticity}}

Neurons typically communicate by action potentials (spikes), transmitter
release and variable spike rates over time. These range from sparse
spiking neurons at 0.1--1~Hz, to medium rates at 10--20~Hz, to
fast-spiking neurons usually in the interneuron domain (40~Hz+). While
neurons vary their baseline rate under excitation or inhibition, the
variance is typically low, about 50--100\% (\protect\hyperlink{ref114}{Scheler 2017}). Neurons also communicate in a
slower fashion by release of neuromodulators or postsynaptic slow
receptors for glutamate and GABA (mGLU receptors, GABA-B).

Another form of neuronal transmission has been explored as a result of
biophysical investigations into membrane excitability (the ``soliton''
theory, (\protect\hyperlink{ref115}{Heimburg 2022})). This would
implement strong and ultrafast direct transmission of excitation across
membranes of neighboring neurons. The transmission is strong enough to
cross the synaptic cleft. This would lead to chains of neurons which can
be activated in this fundamentally different way. Building chains of
neurons (or paths through a network) is in general a good strategy to
ensure fast transmission in spite of recurrence. The soliton type of
transmission may be of particular interest, because, in contrast to
transmitter release/receptor activation, it may not undergo plasticity.
A significant component of neural transmission may be thus devoid of
plasticity.

\hypertarget{discussion}{%
\subsection*{Discussion}\label{discussion}}

\hypertarget{adaptation-as-control-of-an-open-system}{%
\subsubsection*{Adaptation as control of an open
system}\label{adaptation-as-control-of-an-open-system}}

Current theories of synaptic plasticity implement a model of memory
where traces are built from signals at synapses and result in weight
changes. In this way, data are learned or stores in the weights of the
network and patterns can be retrieved and interpolated (generalization).
Memory is also reversible, i.e. can be obliterated by new traces. From a
cognitive and behavioral perspective, memory does not appear as a
passive recording of data input, but as a selection based on an existing
model. It is also not an operation where a blank slate is inscribed and
wiped again, but rather a mechanism where permanent changes occur when
memories are stored, e.g. ``extinction'' as opposed to erasure of fear
events (\protect\hyperlink{ref116}{Myers and Davis 2007}). Furthermore,
a developmental trajectory exists from neonate to juvenile to adult to
aged individual, with effects on processes of memorization and
retrieval. The biological brain memorizes few items of information, much
detail is lost within minutes, hours or overnight. Information that is
sufficiently represented by existing knowledge is often used for
adaptation, other items are often simply discarded, i.e. processed
without leaving a memory trace. In general, much information that is
processed on an everyday basis undergoes abstraction and may become part
of a schema complex that it adds to in small ways.

We outlined a neural model based on external parameters at the membrane,
which are relevant for the transmission of information to other neurons,
and internal parameters, hidden from interactions, which store and
filter information and are capable of adapting external values. The core
system programs and re-programs the internal processing system. The
difference between novel signals and existing information drives
adaptation. It forces a system, such as the brain, to remain an open
system with respect to its environment. Such a system lacks the
properties of a formally 'closed' system with fixed properties. It is
necessary to control such a system, to prevent it from arbitrary
adaptations. Memory management is central for that. It ensures
consistency, promotes signal loss where appropriate (\protect\hyperlink{ref117}{Scheler 2001}), and prevents 'catastrophic
forgetting' (\protect\hyperlink{ref118}{Grossberg 2020};
\protect\hyperlink{ref93}{Paik et al. 2020}), i.e. continuous adaptation
without control.

\hypertarget{internal-memory-drives-selection}{%
\subsubsection*{Internal memory drives
selection}\label{internal-memory-drives-selection}}

A very general perspective of what drives neuronal plasticity is the
difference between stored information and new, incoming signals. There
is a role for dynamic internal states in selecting new information for
storage. The mechanisms that guide the interaction of internal state
parameters and incoming information can be defined along the following
lines:

\begin{enumerate}
\def\labelenumi{\arabic{enumi}.}
\item
  small and transient fluctuations are discounted unless the internal
  state is highly receptive,
\item
  strong signals engage the internal processing system and may signal to
  the core for long-term plasticity,
\item
  the ensemble connectivity structure is variable under incoming
  signals.
\end{enumerate}

\hypertarget{synapse-centric-vs.-neuron-centric}{%
\subsubsection*{Synapse-centric vs.
Neuron-centric}\label{synapse-centric-vs.-neuron-centric}}

Any neural system must contain elements

\begin{itemize}
\item
  with high stability which provide the backbone of the knowledge
  structure and are protected against change
\item
  and elements which flexibly adjust to a current situation.
\end{itemize}

Neuronal types, defined by genetics and as the product of evolutionary
learning, are essentially stable, and this stability may be part of the
background structure. Axonal/dendritic morphology is also a stable
component. It is defined during development and requires core-mediated
programs for alteration. The stability-flexibility trade-off is probably
at the center of the capacity of brains to build structured
representations (\protect\hyperlink{ref118}{Grossberg 2020}).

Neurons as processors operate with graded stability from outside to
inside (\protect\hyperlink{f6}{Figure 6}). This principle seems very
intuitive and sound, but it is unusual in computational theory.
Memorization involves transforming transient information (signals) into
gradually more stable information elements. Concentric abstraction of
temporal depth provides short-term adaptation, long-term stability and
the capacity to store information into increasingly stable formats (\protect\hyperlink{f6}{Figure 6}).

\begin{figure*}[htb]
\begin{center}
\includegraphics[width=0.6\textwidth]{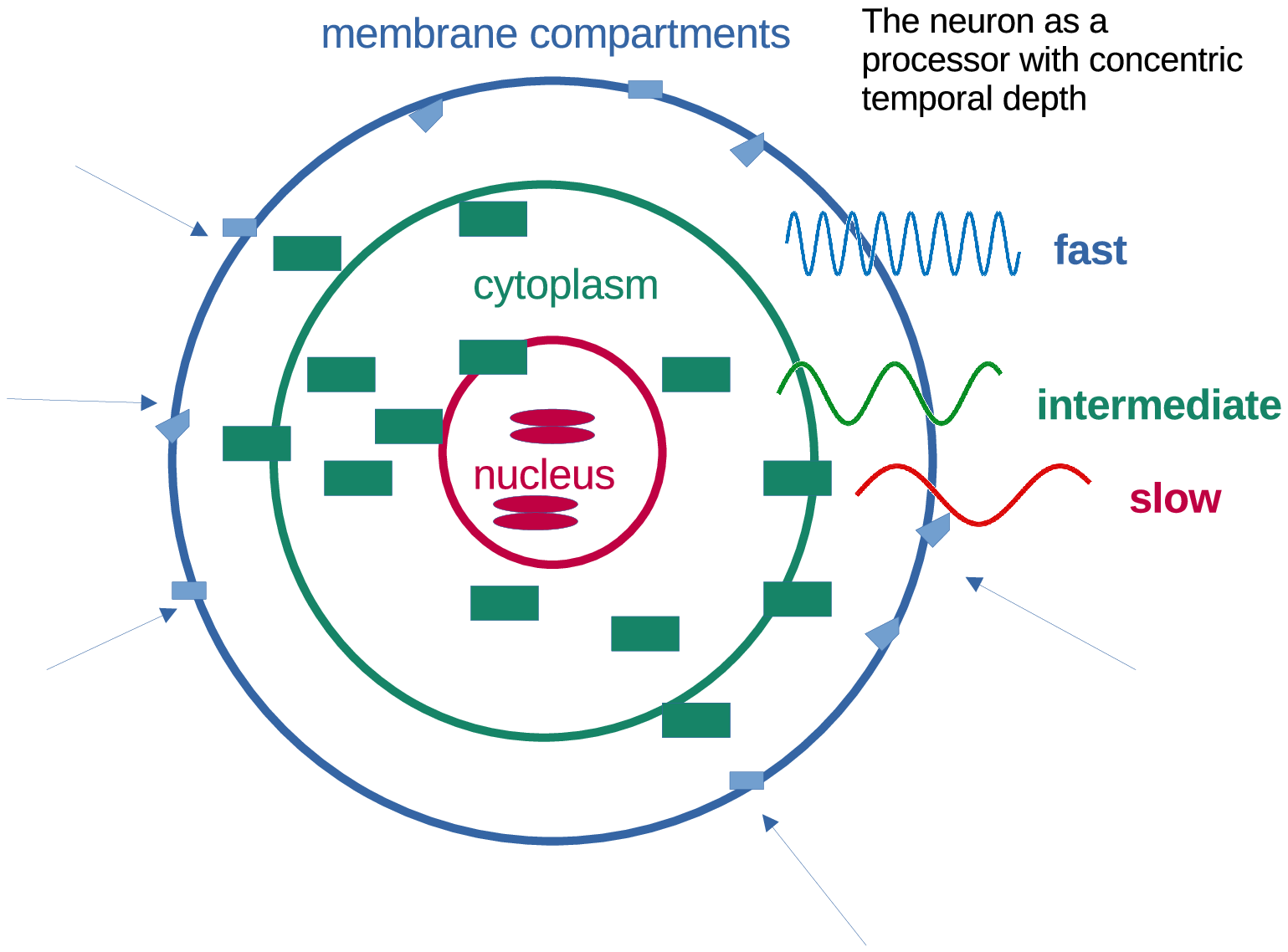}
\end{center}
\caption{The neuron - drawn as a prototypical cell - acts as a processor which receives fast signals and responds with fast (external) and slow (core) adaptations.}
\label{fig:concentric-processor}
\end{figure*}

A synapse-centric theory of memory has a number of unrealistic
consequences:

\begin{itemize}
\item
  synapses are adapted on the basis of external signals. They are
  adapted independently of the neuronal cell state.
\item
  the unit of adaptation is the synapse, not the neuron or the ensemble,
  Global learning rules like backpropagation or credit assignment in
  reinforcement learning attempt to overcome this restriction. They
  focus again on the synapse.
\item
  adaptation in ion channel expression or NM receptor placement are not
  part of a synapse-centric theory.
\item
  since synapses have high turnover (\protect\hyperlink{ref119}{Mongillo et al. 2017};
  \protect\hyperlink{ref120}{Fauth et al. 2015}), information is
  periodically lost or overwritten, unless it is transferred to other
  synapses in a pattern completion process (\protect\hyperlink{ref120}{Fauth et al. 2015}). Memory has to be moved
  just to make up for the biological limitation of the synapse.
\end{itemize}

Another major problem with the synaptic hypothesis is that adjusting
weights in graphs has low representational power. Treating neurons as
adjustable activation functions improves efficiency of implementation,
but does not significantly extend the expressive power of the system (\protect\hyperlink{ref121}{Scheler 2004b};
\protect\hyperlink{ref122}{Kang and Palmer-Brown 2006};
\protect\hyperlink{ref123}{Soltiz et al. 2013}). When such systems store
whole knowledge bases, they acquire and store data sequentially, and
they require enormously large graphs because of a lack of
compressibility. They also overwrite storage by recency, not by
selection. This leads to problems for more complex hierarchical
knowledge and action plans (\protect\hyperlink{ref124}{Hosseini-Asl et
al. 2020}).

In response to some of the challenges of synaptic adaptation models,
numerous improvements and workarounds have been suggested. For instance,
separate memory structures were proposed to solve problems of
sequentiality and compressibility (\protect\hyperlink{ref125}{Graves et
al. 2014}; \protect\hyperlink{ref126}{Hochreiter and Schmidhuber 1997}),
(cf. \protect\hyperlink{f7}{Figure 7}). When memory is conceived as
separate from processing, selection of memory traces becomes possible (\protect\hyperlink{ref156}{Singh and Sutton 1996}).
Another recent improvement (\protect\hyperlink{ref127}{Graves 2014}) involves the insertion of large
amounts of residual connections (``attention'') into learned graphs to
create giant connected networks. This allows for significantly enhanced
pattern completion, even for large and complex patterns (``transformer
architectures''). However, the problems of efficient modularization, of
hierarchical schemata abstraction and inferencing are not addressed by
these models. A system which combines ensemble processing, selection of
items and cascaded depth in internal storage might be able to solve some
of these problems.

\begin{figure*}[htb]
{\bf A}
\begin{minipage}[b]{0.42\textwidth}
\includegraphics[width=1.15\textwidth]{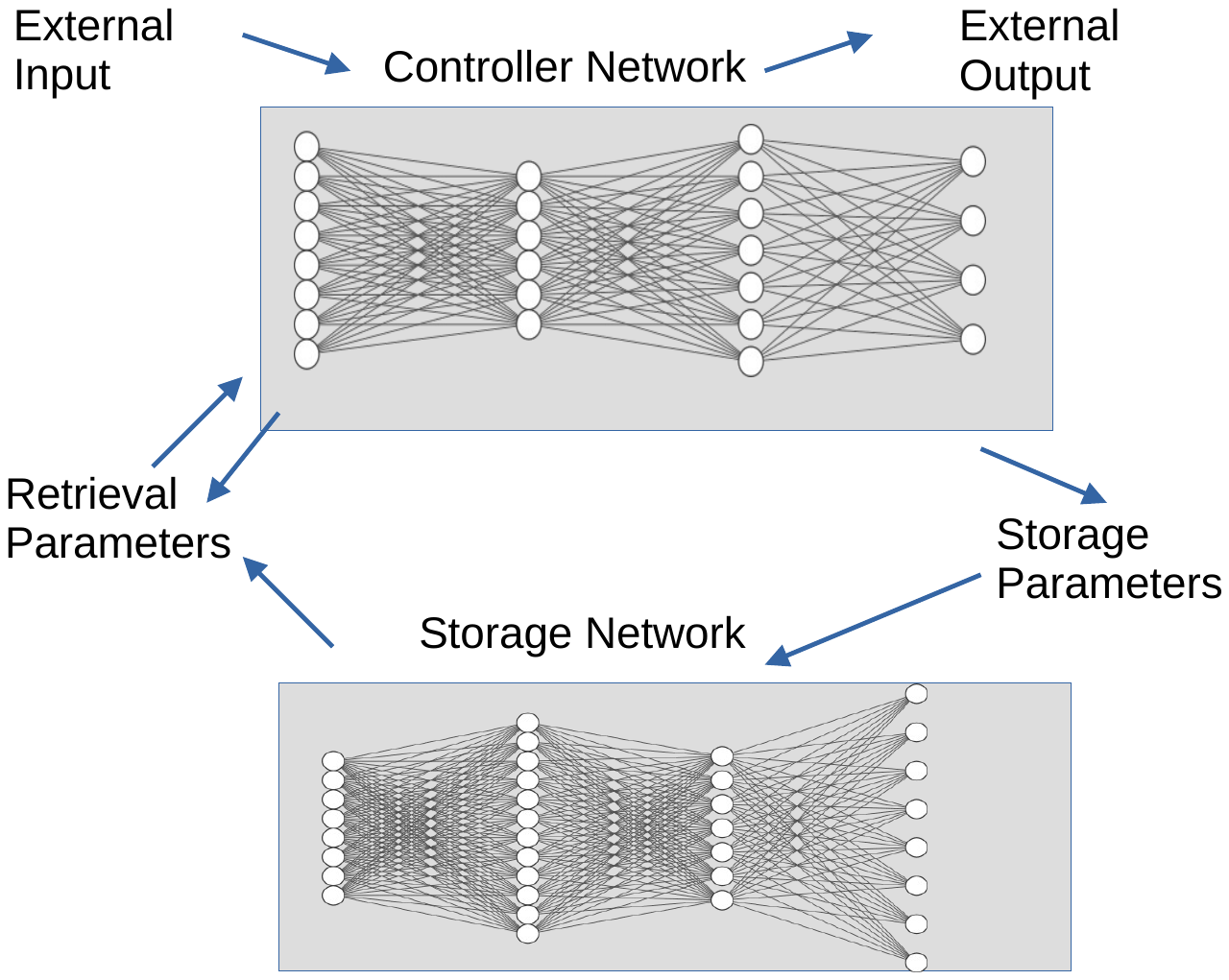}
\end{minipage}
{\bf B}
\begin{minipage}[b]{0.52\textwidth}
\ \vspace*{-3.2cm}\hspace*{-0.6cm}\includegraphics[width=1.6\textwidth]{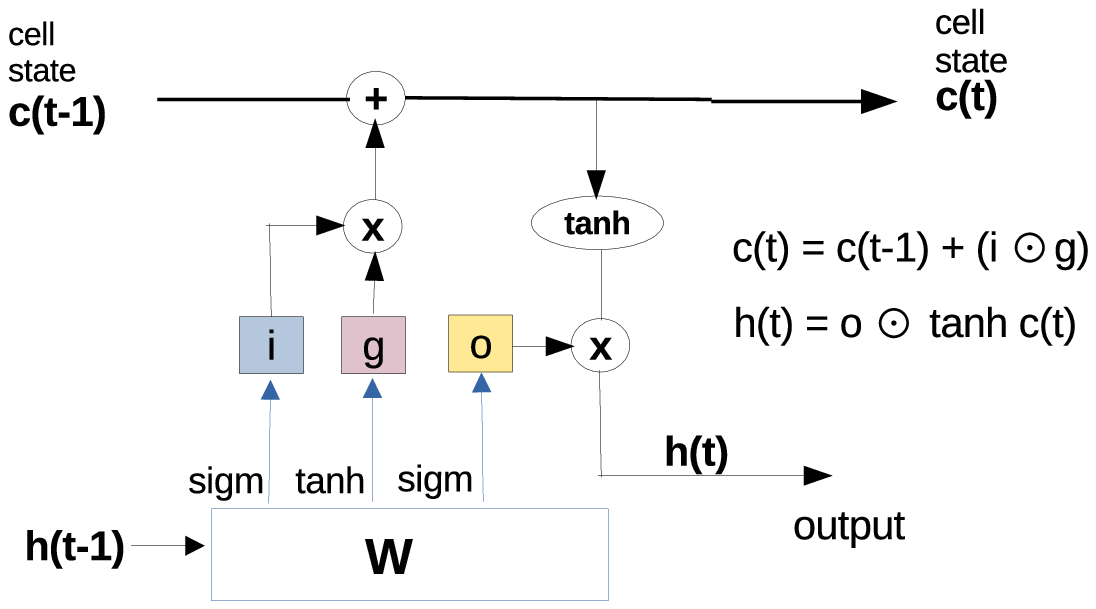}
\end{minipage}
\caption{A. The Neural Turing Machine (NTM) 
(\protect\hyperlink{ref39}{Graves et al. 2014})
is a combination of two recurrent neural networks, one for memory (slow) and one for control/access (fast). B. A long short-term memory (LSTM) cell 
(\protect\hyperlink{ref126}{Hochreiter and Schmidhuber 1997})
uses input (i) and output (o) gates for 'memory cells' (c) and computes a hidden state (h) as an internal storage parameter.}
\label{fig:memory-in-networks}
\end{figure*}

\hypertarget{applications-for-brain-health}{%
\subsubsection*{Applications for brain
health}\label{applications-for-brain-health}}

An important advance of the proposed model is that it provides a
blueprint for a biologically adequate paradigm with applications to both
psychological and medical disease models. There are many extremely
detailed, biologically realistic models, both large (\protect\hyperlink{ref128}{Ito et al. 2026};
\protect\hyperlink{ref129}{Shichkova et al. 2025};
\protect\hyperlink{ref130}{Ali et al. 2026}) and small (\protect\hyperlink{ref20}{Scheler 2013},
\protect\hyperlink{ref73}{2014a}, \protect\hyperlink{ref113}{2018}),
which however lack applicability to even very simple tasks. There
remains an unmet need for brain theories and models that demonstrate
robust functionality. Such frameworks should incorporate the neuronal
cell as a fundamental component, enabling them to adequately capture
disruptions of brain systems that arise from disease processes on
molecular levels. Many diseases of the brain relate to processes,
molecules and interactions that are insufficiently described within the
paradigm of synaptic-centric plasticity.

We want to put forward a single example of a pharmacological effect
which illustrates the limitations of synaptic plasticity models and
expands the view towards neuron-centric models. It was found (\protect\hyperlink{ref131}{Lopez et al. 2022};
\protect\hyperlink{ref132}{Shinohara et al. 2021};
\protect\hyperlink{ref133}{Kang et al. 2022}) that ketamine as a drug
drives genetic expression of the KCNQ2 channel, which increases the IM
(K+) current. In pyramidal cells, e.g. in hippocampus, this reduces
bursting and increases reset after spike firing, limiting the effects of
afterdepolarization (ADP). It also reduces spontaneous glutamate release
at the excitatory synapse. The one-time increase of the genetic
expression of an ion channel provides effects for many weeks. From this
perspective the long-lasting anti-depressant effect of single-dose
ketamine (\textbf{Abdallah et al. 2015}) for psychiatric patients can be
explained and modeled.

This type of result is easy to incorporate into a neuron-centric model
with its integration of neuronal and synaptic properties, and its focus
on levels of persistence in the internal part of the cell (cf.
``write-once'' memory). A synaptic weight adjustment model is too poor
to adequately model such effects -- if such problems are tackled at all,
they are solved by various additions to synaptic weight adjustment as
the expression of all forms of memory (\protect\hyperlink{ref2}{Citri
and Malenka 2008}; \protect\hyperlink{ref136}{Ramirez and Arbuckle
2016}; \protect\hyperlink{ref137}{Magee and Grienberger 2020}),
resulting in ``epicycles'' (arbitrary amendments) of the main paradigm.

Memorization relies on biochemical events. Neuron-centric theories with
internal computation can be considered a substantial way forward,
especially compared to synaptic associative weight adjustment models.
They derive data and results from biophysics, electrophysiology and
molecular biology while the theory itself results in an abstract, and
highly functional model. The proposed neuron model will allow to realize
self-programming of neurons, support modularization into ensembles or
groups, and instantiate complex models of cognition. The brain, of
course, has many additional components, such as glia cells (astrocytes,
oligodendrocytes), the capillary and glymphatic networks, and the
signaling molecules that pass the blood-brain barrier. To reach the goal
of even more comprehensive brain theories we need first to build models
with the full capacity of the neuronal cell.

\hypertarget{conclusion-and-outlook}{%
\subsection*{Conclusion and outlook}\label{conclusion-and-outlook}}

Theoretical neuroscience has focused on synaptic weights as the main
memory mechanism for a considerable time. There are several reasons for
this. The neuron is a highly specialized cell type, strongly
compartmentalized with extensive axons and dendrites. The synapse is a
highly specialized connective element. Several types of neurons have
spiny extrusions of the dendritic membrane with several synapses, spine
shaft elements, and internal proteins and RNA which develop and
disappear within hours or days. The postsynaptic density protein
composition, and the presynaptic vesicle release mechanism are very
intricate cellular structures. But the experimental literature on
neuronal plasticity shows that synapses are an abstract constructive
element, and there are other components of neuronal plasticity (\protect\hyperlink{ref138}{West and Greenberg 2011};
\protect\hyperlink{ref139}{Mahon et al. 2003};
\protect\hyperlink{ref55}{Debanne et al. 2019};
\protect\hyperlink{ref140}{Sehgal et al. 2014};
\protect\hyperlink{ref141}{Nataraj et al. 2010};
\protect\hyperlink{ref142}{Gill and Hansel 2020};
\protect\hyperlink{ref121}{Scheler 2004b},
\protect\hyperlink{ref105}{2014c}; \protect\hyperlink{ref143}{Tully et
al. 2014}). Of particular relevance are plasticity via ion channels and
NM receptors. The internal signaling system constitutes a processing
layer with access both to long-term information in the nucleus and
incoming signals at the membrane from the horizontal network
interactions (\protect\hyperlink{ref138}{West and Greenberg 2011}).
Therefore we focus on the neuron as the relevant unit for plasticity.
Glia-induced and other forms of non-neuronal plasticity are not
discussed here, even though they add to functioning neural ensembles.
The development of a theoretical framework, linking all those aspects of
neural plasticity, is a difficult task. We have begun by re-thinking the
concept of neural plasticity from the perspective of the individual
neuron. We believe that synaptic plasticity is incompletely understood
when placed outside of the context of the individual neuron and its
cellular functioning.

Plasticity has to deal with the problem of stability vs. flexibility,
such as overwriting information over time. There is also the problem of
selecting which information to keep. Since neurons have internal
components which set and re-set responsivity, these problems can be
solved within a neuron-centric framework. We have also put forward a
simple example from disease modeling where the internal dimension of the
neuron and plasticity in protein expression have been shown to be of
significance.

Synaptic weight adaptation models store the history of the use of the
individual synapse, no matter whether by an associative or any other
update rule (such as backpropagation (\protect\hyperlink{ref144}{Rumelhart et al. 1986})). That is very
restrictive. Huge networks with billions of parameters are needed to
solve simple problems by storing millions of patterns. These
restrictions are systematic and mathematically explainable, since the
expressive power of networks with adjustable weights is weak. A
programmable memory at each neuronal site allows for more complex and
concise operations. For instance, patterns can get stored into a small
amount of dedicated neurons (ensembles) for a particular problem. These
ensembles can be activated when similar problems have to be solved, and
by modularization ever new configurations of ensembles to larger
knowledge structures can be constructed. By separating transmission and
storage of information (not every transmission event leaves a trace), we
can focus on the problems of processing and memorization separately, and
achieve a new kind of modular system within the field of large-scale
parallel and distributed computing architectures.

\hypertarget{data-availability}{%
\subsection*{Data availability}\label{data-availability}}

No data are associated with this article.

\cleardoublepage
\onecolumn

\hypertarget{references}{%
\subsection*{References}\label{references}}

\begin{longtable}[]{@{}
  >{\raggedright\arraybackslash}p{(\columnwidth - 2\tabcolsep) * \real{0.0389}}
  >{\raggedright\arraybackslash}p{(\columnwidth - 2\tabcolsep) * \real{0.9611}}@{}}
\toprule\noalign{}
\endhead
\bottomrule\noalign{}
\endlastfoot
\bibno{1} & Abraham WC, Bliss TVP, Collingridge GL, et al.: Long-term
potentiation: 50 years on: past, present and future. \emph{Philos.
Trans. R. Soc. Lond. B Biol. Sci.} 2024;379(6):190620230218. ISSN
0962-8436. 38853569 10.1098/rstb.2023.0218 PMC11343267 \\
\bibno{2} & Abraham WC, Jones OD, Glanzman DL: Is plasticity of synapses the
mechanism of long-term memory storage? \emph{NPJ Sci. Learn.} 2019;4:9.
ISSN 2056-7936. 10.1038/s41539-019-0048-y \\
\bibno{3} & Alejandre-García T, Kim S, Pérez-Ortega J, et al.: Intrinsic
excitability mechanisms of neuronal ensemble formation. \emph{eLife.}
2022;11:5. ISSN 2050-084X. 10.7554/eLife.77470
\url{https://elifesciences.org/articles/77470} \\
\bibno{4} & Ali M, Smiriglia R, Spera E, et al.: Visualization and simulation of
full-scale point-neuron circuits via the neural circuit visualizer web
platform. \emph{Sci. Rep.} 2026 Mar 20;2026. 10.1038/s41598-026-44588-0
\url{https://www.researchgate.net/publication/402960687_Visualization_and_simulation_of_full-scale_point-neuron_circuits_via_the_Neural_Circuit_Visualizer_web_platform} \\
\bibno{5} & Araki Y, Lin D-T, Huganir RL: Plasma membrane insertion of the ampa
receptor glua2 subunit is regulated by nsf binding and q/r editing of
the ion pore. \emph{Proc. Natl. Acad. Sci. USA.}
2010;107(6):11080--11085. ISSN 1091-6490. 10.1073/pnas.1006584107 \\
\bibno{6} & Arshavsky YI: Memory: Axioms and facts. \emph{Neurosci. Behav.
Physiol.} 2021;51(11):1111--1123. ISSN 1573-899X.
10.1007/S11055-021-01171-6 \\
\bibno{7} & Arshavsky YI: Memory: Synaptic or cellular, that is the question.
\emph{Neuroscientist.} 2022;29(5). ISSN 1089--4098.
10.1177/10738584221086488 \\
\bibno{8} & Baudot A, Angelelli J-B, Guénoche A, et al.: Defining a modular
signalling network from the fly interactome. \emph{BMC Syst. Biol.}
2008;2(1):45. ISSN 1752-0509. 10.1186/1752-0509-2-45 \\
\bibno{9} & Baudry M, Zhu G, Liu Y, et al.: Multiple cellular cascades
participate in long-term potentiation and in hippocampus-dependent
learning. \emph{Brain Res.} 2015;1621:73--81. ISSN 1872-6240. 25482663
10.1016/j.brainres.2014.11.033 \\
\bibno{10} & von Bertalanffy L : \emph{General System Theory: Foundations,
Development, Applications}. George Braziller, Inc. revised edition,March
1969. ISBN 0--8076--0453-4. \\
\bibno{11} & Bhalla US: Use of kinetikit and GENESIS for modeling signaling
pathways. \emph{Methods Enzymol.} 2002;345:3--23.
10.1016/s0076-6879(02)45003-3 \\
\bibno{12} & Bhalla US, Iyengar R: Emergent properties of networks of biological
signaling pathways. \emph{Science.} 1999;283:381--387. ISSN 00368075.
10.1126/science.283.5400.381 \\
\bibno{13} & Bloodgood BL, Sabatini BL: Nonlinear regulation of unitary synaptic
signals by cav(2.3) voltage-sensitive calcium channels located in
dendritic spines. \emph{Neuron.} 2007;53(2):249--260.
10.1016/j.neuron.2006.12.017 \\
\bibno{14} & BRAIN Initiative Cell Census Network (BICCN): A multimodal cell
census and atlas of the mammalian primary motor cortex. \emph{Nature.}
Oct 2021;598(7879):86--102. 10.1038/s41586-021-03950-0 \\
\bibno{15} & Bruins Slot LA, Lestienne F, Grevoz-Barret C, et al.: F15063, a
potential antipsychotic with dopamine d(2)/d(3) receptor antagonist and
5-ht(1a) receptor agonist properties: influence on immediate-early gene
expression in rat prefrontal cortex and striatum. \emph{Eur. J.
Pharmacol.} 2009;620(10):27--35. ISSN 1879--0712. 19695244
10.1016/J.EJPHAR.2009.08.019 \\
\bibno{16} & Bulovaite E, Qiu Z, Kratschke M, et al.: A brain atlas of synapse
protein lifetime across the mouse lifespan. \emph{Neuron.}
2022;110(24):4057--4057. ISSN 0896-6273. 10.1016/j.neuron.2022.09.009
\url{https://www.sciencedirect.com/science/article/pii/S0896627322008145} \\
\bibno{17} & Chadi G, Abdallah GS, Duman RS, Krystal JH: Ketamine and
rapid-acting antidepressants: A window into a new neurobiology for mood
disorder therapeutics. \emph{Annu. Rev. Med.} 2015;66:509--523. ISSN
0066-4219. 25341010 10.1146/annurev-med-053013-062946 PMC4428310 \\
\bibno{18} & Chang YM, Jennifer I: Electrophysiological diversity of layer 5
pyramidal cells in the prefrontal cortex of the rhesus monkey: in vitro
slice studies. \emph{J. Neurophysiol.} 2007;98(11):2622--2632. ISSN
0022--3077. 17804579 10.1152/JN.00585.2007 \\
\bibno{19} & Chen HX, Otmakhov N, Lisman J: Requirements for ltp induction by
pairing in hippocampal ca1 pyramidal cells. \emph{J. Neurophysiol.}
1999;82:526--532. ISSN 00223077.
10.1152/JN.1999.82.2.526/ASSET/IMAGES/LARGE/9K0890409005.JPEG \\
\bibno{20} & Ciliberti S, Martin OC, Wagner A: Innovation and robustness in
complex regulatory gene networks. \emph{Proc. Natl. Acad. Sci. USA.}
2007;104(34):13591--13596. ISSN 0027-8424. 10.1073/pnas.0705396104 \\
\bibno{21} & Citri A, Malenka RC: Synaptic plasticity: Multiple forms,
functions, and mechanisms. \emph{Neuropsychopharmacology.}
2008;33(1):18--41. ISSN 0893-133X. 10.1038/sj.npp.1301559 \\
\bibno{22} & Contreras D: Electrophysiological classes of neocortical neurons.
\emph{Neural Networks.} 2004;17(5--6):633--646.
10.1016/j.neunet.2004.04.003 \\
\bibno{23} & Crampin EJ, Schnell S, McSharry PE: Mathematical and computational
techniques to deduce complex biochemical reaction mechanisms.
\emph{Prog. Biophys. Mol. Biol.} 2004;86(9):77--112. ISSN 0079-6107.
\bibno{15261526} 10.1016/j.pbiomolbio.2004.04.002 \\
\bibno{24} & Debanne D: Plasticity of neuronal excitability in vivo. \emph{J.
Physiol.} 2009;587(Pt 13):3057--3058. ISSN 1469-7793. 19567740
10.1113/jphysiol.2009.175448 PMC2727008 \\
\bibno{25} & Debanne D, Inglebert Y, Russier M: Plasticity of intrinsic neuronal
excitability. \emph{Curr. Opin. Neurobiol.} 2019;54(2):73--82. ISSN
09594388. 10.1016/j.conb.2018.09.001
\url{https://linkinghub.elsevier.com/retrieve/pii/S0959438818300709} \\
\bibno{26} & Dehorter N, Ciceri G, Bartolini G, et al.: Tuning of fast-spiking
interneuron properties by an activity-dependent transcriptional switch.
\emph{Science (New York, N.Y.).} 2015;349(9):1216--1220. ISSN 1095-9203.
10.1126/science.aab3415 \\
\bibno{27} & Deng J, Feinberg M, Jones C, et al.: On the steady states of weakly
reversible chemical reaction networks. \emph{arXiv.} 2011;1--20.
10.48550/ARXIV.1111.2386 \\
\bibno{28} & Desai NS: Homeostatic plasticity in the cns: synaptic and intrinsic
forms. \emph{J. Physiol.} 2003;97(4--6):391--402.
10.1016/j.jphysparis.2004.01.005 \\
\bibno{29} & Dolmetsch R: Excitation-transcription coupling: Signaling by ion
channels to the nucleus. \emph{Science's STKE.} 2003;2003(166):pe4. ISSN
2003. 10.1126/stke.2003.166.pe4 \\
\bibno{30} & Donlin-Asp PG, Polisseni C, Klimek R, et al.: Differential
regulation of local mrna dynamics and translation following long-term
potentiation and depression. \emph{Proc. Natl. Acad. Sci. U. S. A.}
2021;118(13):e2017578118. ISSN 10916490. 33771924
10.1073/PNAS.2017578118/-/DCSUPPLEMENTAL \\
\bibno{31} & Fauth M, Wörgötter F, Tetzlaff C: Formation and maintenance of
robust long-term information storage in the presence of synaptic
turnover. \emph{PLoS Comput. Biol.} 2015;11. ISSN 1553-7358. 26713858
10.1371/JOURNAL.PCBI.1004684 \\
\bibno{32} & Fujii H, Bito H: Deciphering Ca 2+-controlled biochemical
computation governing neural circuit dynamics via multiplex imaging.
\emph{Neurosci. Res.} 2022;179:79--90. 10.1016/J.NEURES.2022.04.004 \\
\bibno{33} & Furusawa K, Emoto K: Scrap and build for functional neural
circuits: Spatiotemporal regulation of dendrite degeneration and
regeneration in neural development and disease. \emph{Front. Cell.
Neurosci.} 2021;14--2020. ISSN 1662-5102. 10.3389/fncel.2020.613320 \\
\bibno{34} & Gallistel CR, Matzel LD: The neuroscience of learning: Beyond the
hebbian synapse. \emph{Annu. Rev. Psychol.} 2013;64:169--200. ISSN
0066--4308. 10.1146/annurev-psych-113011-143807 \\
\bibno{35} & Gidon A, Zolnik TA, Fidzinski P, et al.: Dendritic action
potentials and computation in human layer 2/3 cortical neurons.
\emph{Science.} 2020;367:83--87. 31896716 10.1126/science.aax6239 \\
\bibno{36} & Gill DF, Hansel C: Muscarinic modulation of sk2-type k+ channels
promotes intrinsic plasticity in l2/3 pyramidal neurons of the mouse
primary somatosensory cortex. \emph{eNeuro.} 2020;7(2):1--10. ISSN
23732822. 32005752 10.1523/ENEURO.0453-19.2020 PMC7294454 \\
\bibno{37} & Gouwens NW, Sorensen SA, Berg J, et al.: Classification of
electrophysiological and morphological neuron types in the mouse visual
cortex. \emph{Nat. Neurosci.} July 2019;22(7):1182--1195. 31209381
10.1038/s41593-019-0417-0 PMC8078853 \\
\bibno{38} & Graves A: Generating sequences with recurrent neural networks,2014.
\url{https://arxiv.org/abs/1308.0850} \\
\bibno{39} & Graves A, Wayne G, Danihelka I: Neural turing machines.
\emph{arXiv.} 2014;10. \url{http://arxiv.org/abs/1410.5401} \\
\bibno{40} & Grossberg S: A path toward explainable ai and autonomous adaptive
intelligence: Deep learning, adaptive resonance, and models of
perception, emotion, and action. \emph{Front. Neurorobot.} 2020;14. ISSN
1662-5218. 32670045 10.3389/FNBOT.2020.00036 \\
\bibno{41} & Gulledge AT, Kampa BM, Stuart GJ: Synaptic integration in dendritic
trees. \emph{J. Neurobiol.} 2005;64:75--90. ISSN 0022-3034. 15884003
10.1002/NEU.20144 \\
\bibno{42} & Guyon I, Elisseeff A: An introduction to variable and feature
selection. \emph{J. Mach. Learn. Res.} 2003;3:1157--1182.
10.5555/944919.944968 \\
\bibno{43} & Guzowski JF, McNaughton BL, Barnes CA, Worley PF:
Environment-specific expression of the immediate-early gene arc in
hippocampal neuronal ensembles. \emph{Nat. Neurosci.}
1999;2(12):1120--4. ISSN 1097-6256. 10.1038/16046
\url{http://www.nature.com/articles/nn1299_1120} \\
\bibno{44} & Guzowski JF, Setlow B, Wagner EK, et al.: Experience-dependent gene
expression in the rat hippocampus after spatial learning: A comparison
of the immediate-early genes arc, c-fos, and zif268. \emph{J. Neurosci.}
2001;21(14):5089--5098. ISSN 0270--6474. 11438584
10.1523/JNEUROSCI.21-14-05089.2001 PMC6762831 \\
\bibno{45} & Han DH, Park P, Choi DI, et al.: The essence of the engram:
Cellular or synaptic? \emph{Semin. Cell Dev. Biol.} 2022;125:122--135.
ISSN 1084--9521. 10.1016/j.semcdb.2021.05.033
\url{https://www.sciencedirect.com/science/article/pii/S1084952121001488} \\
\bibno{46} & Haney S, Bardwell L, Nie Q: Ultrasensitive responses and
specificity in cell signaling. \emph{BMC Syst. Biol.} 2010;4(1):119.
ISSN 1752-0509. 10.1186/1752-0509-4-119 \\
\bibno{47} & Hayashi T, Huganir RL: Tyrosine phosphorylation and regulation of
the ampa receptor by src family tyrosine kinases. \emph{J. Neurosci.}
Jul 2004;24(27):6152--6160. 15240807 10.1523/JNEUROSCI.0799-04.2004
PMC6729658 \\
\bibno{48} & Heimburg T: The thermodynamic soliton theory of the nervous impulse
and possible medical implications. \emph{Prog. Biophys. Mol. Biol.}
2022;173:24--35. ISSN 0079-6107. 10.1016/j.pbiomolbio.2022.05.007
\url{https://www.sciencedirect.com/science/article/pii/S0079610722000529} \\
\bibno{49} & Hochreiter S, Schmidhuber J: Long short-term memory. \emph{Neural
Comput.} Nov 1997;9(8):1735--1780. ISSN 0899-7667.
10.1162/neco.1997.9.8.1735 \\
\bibno{50} & Hosseini-Asl E, McCann B, Wu C-S, et al.: A simple language model
for task-oriented dialogue. \emph{CoRR.} 2020:abs/2005.00796.
10.48550/ARXIV.2005.00796 \url{https://arxiv.org/abs/2005.00796} \\
\bibno{51} & Humeau Y, Lüthi A: Dendritic calcium spikes induce bi-directional
synaptic plasticity in the lateral amygdala. \emph{Neuropharmacology.}
2007;52(1):234--243. ISSN 0028-3908. 16890250
10.1016/J.NEUROPHARM.2006.07.010 \\
\bibno{52} & Hunter T: Protein kinases and phosphatases: the yin and yang of
protein phosphorylation and signaling. \emph{Cell.} 1995;80(2):225--236.
ISSN 0092-8674. 7834742 10.1016/0092-8674(95)90405-0 \\
\bibno{53} & Ito S, Haufler D, Galván Fraile J, et al.: Deep-learning-assisted
simulation of a cortical circuit: integrating anatomy, physiology and
function. \emph{bioRxiv.} 2026. 10.64898/2026.03.13.711751 \\
\bibno{54} & Jarvis KF: \emph{Computationally Modeling Dynamic Biological
Systems.} University of Maine;2021.
\url{https://books.google.de/books?id=dQX2zgEACAAJ} \\
\bibno{55} & Johenning FW, Theis A-K, Pannasch U, et al.: Ryanodine receptor
activation induces long-term plasticity of spine calcium dynamics.
\emph{PLoS Biol.} 2015;13:e1002181. 26098891
10.1371/journal.pbio.1002181 PMC4476683 \\
\bibno{56} & Josselyn SA, Tonegawa S: Memory engrams: Recalling the past and
imagining the future. \emph{Science. (New York, N.Y.)} 2020;367(6473).
ISSN 10959203. 10.1126/SCIENCE.AAW4325 PMC7577560 \\
\bibno{57} & Kanamori T, Togashi K, Koizumi H, Emoto K: Dendritic remodeling:
Lessons from invertebrate model systems. \emph{Int. Rev. Cell Mol.
Biol.} 2015;318:1--25. Epub 2015 Jun 4. 10.1016/bs.ircmb.2015.05.001 \\
\bibno{58} & Kang M, Palmer-Brown D: A multi-layer adaptive function neural
network (madfunn) for analytical function recognition. \emph{IEEE
International Conference on Neural Networks - Conference Proceedings.}
2006;1784--1789. ISSN 10987576. 10.1109/IJCNN.2006.246895 \\
\bibno{59} & Kang MJY, Hawken E, Vazquez GH: The mechanisms behind rapid
antidepressant effects of ketamine: A systematic review with a focus on
molecular neuroplasticity. \emph{Front. Psychiatry}.2022;13(4): ISSN
1664-0640. 10.3389/FPSYT.2022.860882
\url{https://www.frontiersin.org/articles/10.3389/fpsyt.2022.860882/full} \\
\bibno{60} & Kasai H, Ziv NE, Okazaki H, et al.: Spine dynamics in the brain,
mental disorders and artificial neural networks. \emph{Nat. Rev.
Neurosci.} 2021;22(7):407--422. ISSN 1471-0048. 34050339
10.1038/S41583-021-00467-3 \\
\bibno{61} & Kim EJ, Juavinett AL, Kyubwa EM, et al.: Three types of cortical
layer 5 neurons that differ in brain-wide connectivity and function.
\emph{Neuron.} Dec 2015;88(6):1253--1267. 26671462
10.1016/j.neuron.2015.11.002 PMC4688126 \\
\bibno{62} & Kullmann DM, Lamsa KP: Long-term synaptic plasticity in hippocampal
interneurons. \emph{Nat. Rev. Neurosci.} 2007;8(9):687--699. ISSN
1471-003X. 17704811 10.1038/nrn2207 \\
\bibno{63} & Langille JJ, Gallistel CR: Locating the engram: Should we look for
plastic synapses or information-storing molecules? \emph{Neurobiol.
Learn. Mem.} 2020;169(3):107164. ISSN 10747427.
10.1016/j.nlm.2020.107164 \\
\bibno{64} & Larsen RS, Sjöström PJ: Synapse-type-specific plasticity in local
circuits. \emph{Curr. Opinion Neurobiol.} 35:127--135,2015; ISSN
0959-4388. Circuit plasticity and memory.35:127--135.
10.1016/j.conb.2015.08.001
\href{https://www.sciencedirect.com/science/article/pii/S0959438815001282}{Reference
Source} \\
\bibno{65} & Lee PR, Cohen JE, Iacobas DA, et al.: Gene networks activated by
specific patterns of action potentials in dorsal root ganglia neurons.
\emph{Sci. Rep.} 2017;7:43765. ISSN 2045-2322. 28256583
10.1038/srep43765 \\
\bibno{66} & Lei S, Pelkey K a, Topolnik L, et al.: Depolarization-induced
long-term depression at hippocampal mossy fiber-ca3 pyramidal neuron
synapses. \emph{J. Neurosci.} 2003;23(30):9786--9795. ISSN 1529-2401.
\bibno{14586006} 10.1523/JNEUROSCI.23-30-09786.2003 PMC6740888 \\
\bibno{67} & Li Q, Rothkegel M, Xiao ZC, et al.: Making synapses strong:
Metaplasticity prolongs associativity of long-term memory by switching
synaptic tag mechanisms. \emph{Cerebral Cortex.} 2014;24(2):353--363.
ISSN 1460--2199. 10.1093/cercor/bhs315 \\
\bibno{68} & Lisman J, Schulman H, Cline H: The molecular basis of camkii
function in synaptic and behavioural memory. \emph{Nat. Rev. Neurosci.}
Mar 2002;3(3):175--190. 11994750 10.1038/nrn753 \\
\bibno{69} & Liu S, Anderson PJ, Rajagopal S, et al.: G protein-coupled
receptors: A century of research and discovery. \emph{Circ. Res.}
2024;135(1):174--197. 38900852 10.1161/CIRCRESAHA.124.323067
PMC11192237 \\
\bibno{70} & Lopez JP, Lücken MD, Brivio E, et al.: Ketamine exerts its
sustained antidepressant effects via cell-type-specific regulation of
kcnq2. \emph{Neuron.} 2022;110:2283--2298.e9. ISSN 08966273. 35649415
10.1016/j.neuron.2022.05.001
\url{https://linkinghub.elsevier.com/retrieve/pii/S0896627322004093} \\
\bibno{71} & Magee JC, Grienberger C: Synaptic plasticity forms and functions.
\emph{Annu. Rev. Neurosci.} 2020;43:95--117.
10.1146/annurev-neuro-090919-022842 \\
\bibno{72} & Mahon S, Casassus G, Mulle C, Charpier S: Spike-dependent intrinsic
plasticity increases firing probability in rat striatal neurons in vivo.
\emph{J. Physiol.} 2003;550(Pt 3):947--959. ISSN 0022-3751. 12844508
10.1113/jphysiol.2003.043125 PMC2343063 \\
\bibno{73} & Makino H, Malinow R: Compartmentalized versus global synaptic
plasticity on dendrites controlled by experience. \emph{Neuron.}
2011;72(6):1001--1011. ISSN 0896-6273. 10.1016/j.neuron.2011.09.036
\url{https://www.sciencedirect.com/science/article/pii/S0896627311009937} \\
\bibno{74} & Malenka RC, Bear MF: LTP and LTD: An embarrassment of riches.
\emph{Neuron.} 2004;44(1):5--21. ISSN 08966273. 15450156
10.1016/j.neuron.2004.09.012 \\
\bibno{75} & Manninen T, Hituri K, Kotaleski JH, et al.: Postsynaptic signal
transduction models for long-term potentiation and depression.
\emph{Front. Comput. Neurosci.} 2010;4. ISSN 1662-5188.
10.3389/fncom.2010.00152 \\
\bibno{76} & Mao X, Staiger JF: Multimodal cortical neuronal cell type
classification. \emph{Pflügers Arch.} May 2024;476(5):721--733. 38376567
10.1007/s00424-024-02923-2 PMC11033238 \\
\bibno{77} & Maus L, Lee CK, Altas B, et al.: Ultrastructural correlates of
presynaptic functional heterogeneity in hippocampal synapses. \emph{Cell
Rep.} 2020;30(3):3632--3643.E8. ISSN 22111247.
10.1016/j.celrep.2020.02.083
\url{https://linkinghub.elsevier.com/retrieve/pii/S2211124720302576} \\
\bibno{78} & McFarlan A, Chou C, Watanabe A, et al.: The plasticitome of
cortical interneurons. \emph{Nat. Rev. Neurosci.} 2022;24:80--97.
10.1038/s41583-022-00663-9 \\
\bibno{79} & Mellström B, Savignac M, et al.: Ca 2+ operated transcriptional
networks: Molecular mechanisms and in vivo models. \emph{Physiol. Rev.}
2008;884:421--449. ISSN 0031-9333. 10.1152/physrev.00041.2005 \\
\bibno{80} & Minatohara K, Akiyoshi M, Okuno H: Role of immediate-early genes in
synaptic plasticity and neuronal ensembles underlying the memory trace.
\emph{Front. Mol. Neurosci.} 2016;8. ISSN 16625099. 26778955
10.3389/fnmol.2015.00078 PMC4700275 \\
\bibno{81} & Mongillo G, Rumpel S, Loewenstein Y: Intrinsic volatility of
synaptic connections --- a challenge to the synaptic trace theory of
memory. \emph{Curr. Opin. Neurobiol.} 2017;46(10):7--13. ISSN 09594388.
10.1016/j.conb.2017.06.006
\url{https://linkinghub.elsevier.com/retrieve/pii/S0959438817300673} \\
\bibno{82} & Myers KM, Davis M: Mechanisms of fear extinction. \emph{Mol.
Psychiatry.} 2007;12(3):120--150. 10.1038/sj.mp.4001939 \\
\bibno{83} & Nataraj K, Le Roux N, Nahmani M, et al.: Visual deprivation
suppresses l5 pyramidal neuron excitability by preventing the induction
of intrinsic plasticity. \emph{Neuron.} 2010;68(4):750--762. ISSN
1097-4199. 21092863 10.1016/j.neuron.2010.09.033 PMC2990987 \\
\bibno{84} & Nimchinsky EA, Sabatini BL, Svoboda K: Structure and function of
dendritic spines. \emph{Annu. Rev. Physiol.} 2002;64:313--353. ISSN
0066-4278. 11826272 10.1146/ANNUREV.PHYSIOL.64.081501.160008 \\
\bibno{85} & Nishiyama M, Hong K, Mikoshiba K, et al.: Calcium stores regulate
the polarity and input specificity of synaptic modification.
\emph{Nature.} 2000;408(6812):584--588. 11117745 10.1038/35046067 \\
\bibno{86} & O'Leary T, Williams AH, Franci A, et al.: Cell types, network
homeostasis, and pathological compensation from a biologically plausible
ion channel expression model. \emph{Neuron.} 2014;82(4):809--821. ISSN
08966273. 24853940 10.1016/j.neuron.2014.04.002 PMC4109293
\url{https://linkinghub.elsevier.com/retrieve/pii/S089662731400292X} \\
\bibno{87} & Özçete ÖD, Banerjee A, Kaeser PS: Mechanisms of neuromodulatory
volume transmission. \emph{Mol. Psychiatry.} 2024;29:3680--3693.
\bibno{38789677} 10.1038/s41380-024-02608-3 PMC11540752 \\
\bibno{88} & Paik I, Oh S, Kwak T, Kim I: Overcoming catastrophic forgetting by
neuron-level plasticity control. \emph{Proc. AAAI Conf. Artif. Intell.}
2020;34(4):5339--5346. ISSN 2374-3468. 10.1609/aaai.v34i04.5981
\url{https://aaai.org/ojs/index.php/AAAI/article/view/5981} \\
\bibno{89} & Park A, Jacob AD, Walters BJ, et al.: A time-dependent role for the
transcription factor creb in neuronal allocation to an engram underlying
a fear memory revealed using a novel in vivo optogenetic tool to
modulate creb function. \emph{Neuropsychopharmacology.}
2020;45(6):916--924. ISSN 1740-634X. 31837649 10.1038/s41386-019-0588-0
PMC7162924 \\
\bibno{90} & Planert H, Mittermaier F, Grosser S, et al.: Electrophysiological
classification of human layer 2--3 pyramidal neurons reveals
subtype-specific synaptic interactions. \emph{Nat. Neurosci.}
2025;29(2):455--466. 41372679 10.1038/s41593-025-02134-7 PMC12880919 \\
\bibno{91} & Quiroga RQ: Concept cells: the building blocks of declarative
memory functions. \emph{Nat. Rev. Neurosci.} 2012;13(8):587--597.
10.1038/nrn3242 \\
\bibno{92} & Ramaswamy S, Markram H: Anatomy and physiology of the thick-tufted
layer 5 pyramidal neuron. \emph{Front. Cell. Neurosci.} 2015;9:1--29.
ISSN 1662-5102. 10.3389/fncel.2015.00233
\url{http://journal.frontiersin.org/Article/10.3389/fncel.2015.00233/abstract} \\
\bibno{93} & Ramirez A, Arbuckle MR: Synaptic plasticity: the role of learning
and unlearning in addiction and beyond. \emph{Biol. Psychiatry.}
2016;80(9):e73--e75. 27697156 10.1016/j.biopsych.2016.09.002
PMC5347979 \\
\bibno{94} & Redondo R, Morris R: Making memories last: the synaptic tagging and
capture hypothesis. \emph{Nat. Rev. Neurosci.} 2011;12(1):17--30. ISSN
1471-003X. 10.1038/nrn2963
\url{http://www.nature.com/articles/nrn2963} \\
\bibno{95} & Remy S, Spruston N: Dendritic spikes induce single-burst long-term
potentiation. \emph{Proc. Natl. Acad. Sci. USA.} 2007;104:17192--17197.
ISSN 0027-8424. 10.1073/pnas.0707919104 \\
\bibno{96} & Renner J, Rasia-Filho AA: Morphological features of human dendritic
spines. Adv. Neurobiol. 2023;34:367--496.
10.1007/978-3-031-36159-3\_9 \\
\bibno{97} & Rienecker KDA, Poston RG, Segales JS, et al.: Mild membrane
depolarization in neurons induces immediate early gene transcription and
acutely subdues responses to successive stimulus. \emph{J. Biol. Chem.}
2022;298(9):102278. ISSN 0021--9258. 10.1016/J.JBC.2022.102278
\url{http://www.jbc.org/article/S0021925822007207/fulltext} \\
\bibno{98} & Robbins MJ, Critchlow HM, Lloyd A, et al.: Differential expression
of ieg mrna in rat brain following acute treatment with clozapine or
haloperidol: a semi-quantitative rt-pcr study. \emph{J.
Psychopharmacol.} 2008;22(5):536--542. ISSN 0269-8811. 18208916
10.1177/0269881107081521 \\
\bibno{99} & Rockman H, Lefkowitz R: G protein--coupled receptors: from
radioligand binding to cellular signaling. \emph{J. Clin. Invest.}
2024;134:e178109. 38426490 10.1172/JCI178109 PMC10904040 \\
\bibno{100} & Rodrigues YE, Tigaret CM, Marie H, et al.: A stochastic model of
hippocampal synaptic plasticity with geometrical readout of enzyme
dynamics. \emph{eLife.} Aug 2023;12:e80152. ISSN 2050-084X. 37589251
10.7554/eLife.80152 PMC10435238 \\
\bibno{101} & Rogerson T, Cai DJ, Frank A, et al.: Synaptic tagging during
memory allocation. \emph{Nat. Rev. Neurosci.} 2014;15(3):157--169.
\bibno{24496410} 10.1038/nrn3667 PMC3992944 \\
\bibno{102} & Rumelhart DE, McClelland JL; PDP Research Group: \emph{Parallel
Distributed Processing: Explorations in the Microstructure of Cognition.
1: Foundations} . Cambridge, MA: MIT Press;1986. ISBN 026268053X. \\
\bibno{103} & Saak V. O: The dark matter of the brain. \emph{Brain Struct.
Funct.} 2019;224(8):2811--2826. 10.1007/s00429-019-01835-7 \\
\bibno{104} & Sabatini BL, Oertner TG, Svoboda K: The life cycle of ca2+ ions in
dendritic spines. \emph{Neuron.} 2002;33(3):439--452. ISSN 08966273.
\bibno{11832230} 10.1016/S0896-6273(02)00573-1 \\
\bibno{105} & Scheler G: Antagonistic feedforward interactions: Unifying
principle in biological regulation. \emph{ECCS14 (European Conference on
Complex Systems).} Sep 2014a. \\
\bibno{106} & Scheler G: Dynamic re-wiring of protein interaction: The case of
transactivation. \emph{arXiv:q-bio/0609014.} Oct 2006.
10.48550/ARXIV.Q-BIO/0609014 \\
\bibno{107} & Scheler G: Extracellular-to-intracellular signal transfer via
g-proteins. \emph{arXiv:abs/q-bio/0503031.} 2005.
10.48550/ARXIV.Q-BIO/0503031
\url{https://arxiv.org/abs/q-bio/0503031} \\
\bibno{108} & Scheler G: Learning intrinsic excitability in medium spiny neurons
{[}version 2; peer review: 2 approved{]}. \emph{F1000Research.}
2014b;2:88. 25520776 10.12688/f1000research.2-88.v2 PMC4264637 \\
\bibno{109} & Scheler G: Logarithmic distributions prove that intrinsic learning
is hebbian {[}version 2; peer review: 2 approved{]}. \emph{F1000Res.}
2017;6(1222). 10.12688/f1000research.12130.2 \\
\bibno{110} & Scheler G: Memorization in a neural network with adjustable
transfer function and conditional gating. \emph{arXiv:q-bio/0403011.}
2004b;3. 10.48550/arxiv.q-bio/0403011
\url{https://arxiv.org/abs/q-bio/0403011v1} \\
\bibno{111} & Scheler G: Negative feedback is a subordinated control structure
in signal transduction. \emph{F1000Posters.} 2012;3(poster):398.
\url{https://f1000research.com/posters/1090221} \\
\bibno{112} & Scheler G: Neuromodulation influences synchronization and
intrinsic read-out. \emph{F1000Res.} 2018;7(1277):8.
10.12688/f1000research.15804.2 \\
\bibno{113} & Scheler G: Regulation of neuromodulator receptor
efficacy---implications for whole-neuron and synaptic plasticity.
\emph{Prog. Neurobiol.} 2004a;72(6):399--415. ISSN 0301-0082.
10.1016/j.pneurobio.2004.03.008
\url{https://www.sciencedirect.com/science/article/pii/S0301008204000450} \\
\bibno{114} & Scheler G: Signal loss with neural controllers.
\emph{ResearchGate.} 2001;3:2412775.
\url{https://www.researchgate.net/publication/2412775_Signal_Loss_With_Neural_Controllers} \\
\bibno{115} & Scheler G: Timing in cells: invariance and nonsynchronicity.
\emph{F1000Posters.} 5:292 (poster),2014c.
\url{https://f1000research.com/posters/1095459} \\
\bibno{116} & Scheler G: Transfer functions for protein signal transduction:
Application to a model of striatal neural plasticity. \emph{PLoS One.}
2013;8(2):e55762. 23405211 10.1371/journal.pone.0055762 PMC3565992 \\
\bibno{117} & Scheler G, Schumann J: Presynaptic modulation as fast synaptic
switching: state-dependent modulation of task performance.
\emph{Proceedings of the International Joint Conference on Neural
Networks.} 2003;1:218--223. 10.1109/IJCNN.2003.1223347 \\
\bibno{118} & Scheler G, Schumann J: Localist plasticity identified by mutual
information. \emph{bioRxiv.} 2019;658153. 10.1101/658153
\url{https://www.biorxiv.org/content/early/2019/06/03/658153} \\
\bibno{119} & Scheler G, Schumann ML, Schumann J: Localist neural plasticity
identified by mutual information. \emph{Journal of Computational
Neuroscience.} 2025;53:321--331. 40120001 10.1007/s10827-025-00901-w \\
\bibno{154} & Scheler G, Schumann ML, Schumann J: Non-standard memory models with indexed retrieval. \em{arXiv}. 2026. arXiv:2608.27479 [q-bio-NC].\\
\bibno{120} & Schirò G, Iacono S, Ragonese P, et al.: A brief overview on
bdnf-trk pathway in the nervous system: A potential biomarker or
possible target in treatment of multiple sclerosis? \emph{Front.
Neurol.} 2022;13. ISSN 1664-2295. 10.3389/fneur.2022.917527
\href{https://www.frontiersin.org/journals/neurology/articles/Front\%20Neurol.}{Reference
Source} \\
\bibno{121} & Sehgal M, Ehlers VL, Moyer JR: Learning enhances intrinsic
excitability in a subset of lateral amygdala neurons. \emph{Learn. Mem.}
NY: Cold Spring Harbor;2014;21(3):161--170. ISSN 1549-5485. 24554670
10.1101/lm.032730.113 PMC3929854 \\
\bibno{122} & Sharma S, Vinken K, Jagadeesh AV, et al.: Face cells encode object
parts more than facial configuration of illusory faces. \emph{Nat.
Commun.} 2024;15:9879. 39543127 10.1038/s41467-024-54323-w
PMC11564726 \\
\bibno{123} & Shenoy SK, Lefkowitz RJ: $\beta$-arrestin-mediated receptor trafficking
and signal transduction. \emph{Trends Pharmacol. Sci.} 2011;32(9):521.
ISSN 01656147. 10.1016/J.TIPS.2011.05.002 PMC3159699 \\
\bibno{124} & Sherman SM, Usrey WM: A reconsideration of the core and matrix
classification of thalamocortical projections. \emph{J. Neurosci.}
2024b;44(24):e0163242024. 38866538 10.1523/JNEUROSCI.0163-24.2024 \\
\bibno{125} & Sherman SM, Usrey WM: Transthalamic pathways for cortical
function. \emph{J. Neurosci.} 2024a;44(35):e0909242024. 39197951
10.1523/JNEUROSCI.0909-24.2024 \\
\bibno{126} & Shichkova P, Coggan JS, Kanari L, et al.: Breakdown and repair of
metabolism in the aging brain. \emph{Front. Sci.} 2025;3--2025. ISSN
2813-6330. 10.3389/fsci.2025.1441297 \\
\bibno{127} & Shinar G, Feinberg M: Structural sources of robustness in
biochemical reaction networks. \emph{Science (New York, N.Y.)}
2010;327(3):1389--1391. ISSN 1095-9203. 20223989
10.1126/science.1183372 \\
\bibno{128} & Shinohara R, Aghajanian GK, Abdallah CG: Neurobiology of the
rapid-acting antidepressant effects of ketamine: Impact and
opportunities. \emph{Biol. Psychiatry.} 90:85--95,72021. ISSN 18732402.
10.1016/J.BIOPSYCH.2020.12.006 \\
\bibno{129} & Soltiz M, Kudithipudi D, Merkel C, et al.: Memristor-based neural
logic blocks for nonlinearly separable functions. \emph{IEEE Trans.
Comput.} 2013;62(8):1597--1606. 10.1109/TC.2013.75 \\
\bibno{130} & Steward O, Schuman EM: Compartmentalized synthesis and degradation
of proteins in neurons. \emph{Neuron.} 2003;40(10):347--359. ISSN
08966273. 10.1016/S0896-6273(03)00635-4
\url{https://linkinghub.elsevier.com/retrieve/pii/S0896627303006354} \\
\bibno{131} & Traunmüller L, Duffy EE, Liu H, et al.: Novel environment exposure
drives temporally defined and region-specific chromatin accessibility
and gene expression changes in the hippocampus. \emph{Nat. Commun.}
August 2025;16(1):7787. 40841540 10.1038/s41467-025-63029-6
PMC12370903 \\
\bibno{132} & Trettenbrein PC: The demise of the synapse as the locus of memory:
A looming paradigm shift? \emph{Front. Syst. Neurosci.} 2016;10:88. ISSN
1662-5137. 27909400 10.3389/fnsys.2016.00088. PMC5112247 \\
\bibno{133} & Tully PJ, Hennig MH, Lansner A: Synaptic and nonsynaptic
plasticity approximating probabilistic inference. \emph{Front. Synaptic
Neurosci.} 2014 ;8. ISSN 1663--3563. 24782758 10.3389/fnsyn.2014.00008
PMC3986567
\url{http://journal.frontiersin.org/article/10.3389/fnsyn.2014.00008/abstract} \\
\bibno{134} & Usrey WM, Sherman SM: Corticofugal circuits: Communication lines
from the cortex to the rest of the brain. \emph{J. Comput. Neurosci.}
2019;527(3):640--650. 29524229 10.1002/cne.24423 PMC6131091 \\
\bibno{135} & Vayttaden SJ, Bhalla US: Developing complex signaling models using
genesis/kinetikit. \emph{Science's STKE.} 2004;2004(219):pl4.
10.1126/stke.2192004pl4 \\
\bibno{136} & Waidmann EN, Koyano KW, Hong JJ, et al.: Local features drive
identity responses in macaque anterior face patches. \emph{Nat. Commun.}
2022;13:5592. 36151142 10.1038/s41467-022-33240-w PMC9508131 \\
\bibno{137} & West AE, Greenberg ME: Neuronal activity-regulated gene
transcription in synapse development and cognitive function. \emph{Cold
Spring Harb. Perspect. Biol.} 2011;3(6):1--21. ISSN 19430264. 21555405
10.1101/cshperspect.a005744 \\
\bibno{138} & Winnubst J, Lohmann C: Synaptic clustering during development and
learning: The why, when, and how. \emph{Front. Mol. Neurosci.} 2012;5.
ISSN 16625099. 22666187 10.3389/fnmol.2012.00070 PMC3364493 \\
\bibno{139} & Wu J, Rowan MJ, Anwyl R: Long-term potentiation is mediated by
multiple kinase cascades involving camkii or either pka or p42/44 mapk
in the adult rat dentate gyrus in vitro. \emph{J. Neurophysiol.} Jun
2006;95(6):3519--3527. 10.1152/jn.01235.2005 \\
\bibno{140} & Wu Y, Maass W: A simple model for behavioral time scale synaptic
plasticity (btsp) provides content addressable memory with binary
synapses and one-shot learning. \emph{Nat. Commun.} 2025;16:342.
\bibno{39747916} 10.1038/s41467-024-55563-6 PMC11695864 \\
\bibno{147} & Xu B, Shen H, Yu Y: The write-cost bottleneck: energy constraints on continual learning in brains and machines. \emph{Cogn Neurodyn.} 2026 Dec;20(1):139. 10.1007/s11571-026-10508-1.\\
\bibno{141} & Zhong WX, Dong ZF, Tian M, et al.: N-methyl-d-aspartate
receptor-dependent long-term potentiation in ca1 region affects synaptic
expression of glutamate receptor subunits and associated proteins in the
whole hippocampus. \emph{Neuroscience.} 2006;141(1):1399--1413. ISSN
03064522. 10.1016/j.neuroscience.2006.04.070
\url{https://linkinghub.elsevier.com/retrieve/pii/S0306452206006154} \\
\bibno{144} & Zhou Z, Walsh CA: Somatic mosaicism in the brain: linking development, ageing and neurodegeneration. \emph{Nat Rev Neurol.} 2026 Sep 14. 10.1038/s41582-026-01268-x.\\
\bibno{142} & Zhou Y, Won J, Karlsson MG, et al.: Creb regulates excitability
and the allocation of memory to subsets of neurons in the amygdala.
\emph{Nat. Neurosci.} 2009;12(11):1438--43. ISSN 1546--1726.
10.1038/nn.2405 \\
\bibno{143} & Zucker RS, Regehr WG: Short-term synaptic plasticity. \emph{Annu.
Rev. Physiol.} 2002;64:355--405.
10.1146/annurev.physiol.64.092501.114547 \\
\end{longtable}

\end{document}